\documentclass[12pt,preprint]{aastex}
\begin{document}
\def\nuc#1{${}^{#1}$}
\def\nucm#1#2{{}^{#1}{\rm #2}}
\def\feoh{[{\rm Fe}/{\rm H}]}
\newcommand{\rsun}[1]{\ensuremath{{#1} R_{\odot}}}
\newcommand{\alga}{\ensuremath{(\alpha,\gamma)}}
\newcommand{\aln}{\ensuremath{(\alpha, n)}}
\newcommand{\pga}{\ensuremath{(p,\gamma)}}
\newcommand{\bdecp}{\ensuremath{(e^{-},\nu^{+} \gamma)}}
\newcommand{\agco}{$\nucm{12}{C} \alga \nucm{16}{O}$ }
\newcommand{\pgac}{$\nucm{12}{C} \pga \nucm{13}{N}$ }
\newcommand{\pgan}{$\nucm{14}{N} \pga \nucm{15}{O}$ }
\newcommand{\pgao}{$\nucm{16}{O} \pga \nucm{17}{F}$ }
\newcommand{\cncycle}{$\nucm{12}{C} \pga \nucm{13}{N}
 \bdecp \nucm{13}{C} \pga \nuc{14}{N}m }
\newcommand{\ndrip}{$\nucm{13}{C} \aln \nucm{16}{O}$}
\newcommand{\lhyd}{\ensuremath{L_{\textrm{H}}}}
\newcommand{\lhe}{\ensuremath{L_{\textrm{He}}}}
\newcommand{\pow}[2]{\ensuremath{{#1} \times 10^{#2}}}
\newcommand{\powk}[2]{\ensuremath{{#1} \times 10^{#2} \textrm{K}}}
\newcommand{\hb}{\hfill\break}
\newcommand{\refeq}[1]{eq.(\ref{#1})}
\def\mmps{HE0107-5240}
\def\myrm1{$M_\odot \ {\rm yr}^-1$}
\def\msun{M_\odot}
\def\dotc{\dot{\rm C}}
\def\doto{\dot{\rm O}}
\def\coopr{\dot{\rm C}/\dot{\rm O}}

\shortauthors{Suda et al.}
\shorttitle{Is HE0107-5240 A Primordial Star?}

\title{Is HE0107-5240 A Primordial Star? \\
--- The Characteristics of Extremely Metal-Poor Carbon-Rich Stars ---}
\author{Takuma Suda\altaffilmark{1}, Masayuki Aikawa\altaffilmark{1},
Masahiro N. Machida\altaffilmark{2,3}, Masayuki Y. Fujimoto}
\affil{Department of Physics, Hokkaido University, Sapporo 060-0810,
 Japan} 
\email{suda@astro1.sci.hokudai.ac.jp, aikawa@nucl.sci.hokudai.ac.jp,
machida@th.nao.ac.jp, fujimoto@astro1.sci.hokudai.ac.jp }
\and
\author{ Icko, Iben, Jr. \altaffilmark{4}}
\affil{Departments of Astronomy and of Physics, University of
Illinois at Urbana-Champaign, IL 61801}
\email{icko@astro.uiuc.edu}

\altaffiltext{1}{Meme Media Laboratory, Hokkaido University }
\altaffiltext{2}{present address; Center for Frontier Science, 
Chiba University, Yayoicho 1-33, Inageku, Chiba, 263-8522, Japan}
\altaffiltext{3}{National Astronomical Observatory of Japan, Mitaka,
Tokyo, 181-8588, Japan}
\altaffiltext{4}{Visiting JSPS Eminent Professor, Hokkaido University}

\begin{abstract}
We discuss the origin of HE0107-5240 which, with a metallicity
of $ \feoh =-5.3$, is the most iron-poor star yet observed.   
Its discovery has an important bearing on the question of the observability of first generation stars in our Universe.
     In common with other stars of very small metallicity
($-4 \lesssim \feoh \lesssim - 2.5$), HE0107-5240 shows a peculiar
abundance pattern, including large enhancements of C, N, and O,
and a more modest enhancement of Na.
     The observed abundance pattern can be explained by
nucleosynthesis and mass transfer in a first generation binary
star, which, after birth, accretes matter from a primordial
cloud mixed with the ejectum of a supernova.  
     We elaborate the binary scenario on the basis of our current
understanding of the evolution and nucleosynthesis of extremely
metal-poor, low-mass model stars and discuss the possibility of
discriminating this scenario from others.

In our picture, iron-peak elements arise in surface layers of
the component stars by accretion of gas from the polluted primordial
cloud, pollution occurring after the birth of the binary.  
     To explain the observed C, N, O, and Na enhancements as
well as the $\nucm{12}{C}/\nucm{13}{C}$ ratio, we suppose that
the currently observed star, once the secondary in a binary,
accreted matter from a chemically evolved companion, which is
now a white dwarf.  
     To estimate the abundances in the matter transferred in the
binary, we rely on the results of computations of model stars
constructed with up-to-date input physics.   
Nucleosynthesis in a helium flash driven convective zone into which hydrogen has been injected is followed, allowing us to to explain the origin in the primary of the observed O and Na enrichments and to discuss the abundances of s-process elements.

     From the observed abundances, we conclude that \mmps\ has
evolved from a wide binary (of initial separation $\sim 20$ AU)
with a primary of initial mass in the range $1.2 \sim 3 \msun$.
     On the assumption that the system now consists of a white dwarf and a red giant, the present binary separation and period are estimated at $\simeq 34$ AU and a period of $\simeq 150$ years, respectively.

     We also conclude that the abundance distribution of heavy s-process elements may hold the key to a satisfactory understanding of the origin of HE0107-5240.  
     An enhancement of $[{\rm Pb} / {\rm Fe}] \simeq 1 \sim 2$
should be observed if \mmps\ is a second generation star, formed
from gas already polluted with iron-group elements.  
If the enhancement of main-line s-process elements is not detected,
\mmps\ may be a first generation secondary in a binary system
with a primary of mass less than 2.5 $\msun$, born from gas of
primordial composition, produced in the Big Bang, and subsequently subjected to surface
pollution by accretion of gas from the parent cloud metal-enriched
by mixing with the ejectum of a supernova.

\end{abstract}


\section{Introduction}
The discovery of the exceedingly metal-poor red giant \mmps\
\citep{chr02} has great importance for our understanding
of early star formation in the Galaxy. It has the smallest metallicity
($\feoh = -5.3$) of any star yet observed and it exhibits several
abundance ratios which differ markedly from abundance ratios in
solar system material. In particular, it shows large enhancements
of carbon ($[\nucm{}{C}/\nucm{}{Fe}] = 4.0$), nitrogen
($[\nucm{}{N}/\nucm{}{Fe}] = 2.3$),
and oxygen ($\mbox{[O/Fe]}=2.4^{+0.2}_{-0.4}$) \citep{bes04},
as well as a mild enhancement of sodium
($[\nucm{}{Na}/\nucm{}{Fe}] = 0.8$); currently, only an upper
bound exists for the important s-process element Ba 
($[\nucm{}{Ba}/\nucm{}{Fe}] < 0.8$) \citep{chr02}. We use the
traditional term ``metal poor'' in referring to \mmps\ even
though, thanks to the large abundances of CNO elements relative to
Fe, it is certainly not ``heavy element poor.''

Recent papers which discuss the origin of \mmps\ use arguments
involving supernova nucleosynthesis in a primordial cloud
\citep{ume03,lim03}, external pollution after birth
\citep{shi03}, and the mass distribution predicted by theories
of star formation \citep{sal03, omu03}.  
      \citet{shi03} predict a metallicity distribution function
for first generation (Pop.~III) stars currently burning hydrogen
and conclude that \mmps \ is a first generation object with a
surface affected by accreting interstellar matter polluted with
heavy elements.  
\citet{ume03} adjust parameters in a first generation supernova
model in such a way as to produce a C/Fe ratio in the supernova
ejectum that agrees with the ratio observed in \mmps\ and argue
that \mmps\ is a second generation object formed from the
primordial cloud after it has been mixed with the ejectum of
the supernova.
     A similar scenario is presented by \citet{lim03} who argue
that the \mmps\ abundances can be produced by a combination of
two types of supernova ejecta: a normal ejectum consisting of
$\sim 0.06 \msun$ of iron and an abnormal ejectum consisting
only of products of partial helium burning.   
Though all extant scenarios address important aspects of the problem,
further discussion is warranted of the physics of star formation
and of the chemical composition expected in a primordial cloud
into which matter ejected by a supernova has been mixed.  
     More importantly, the modifications of surface abundances
which \mmps\ has suffered during its long life should be elucidated.
In particular, the possibility of accretion from an evolved first
generation companion which has mixed to its surface products of
internal nucleosynthesis should be explored. In this paper, we describe
results of such an exploration.

A major characteristic of models of extremely metal poor stars is
that, although the p-p chain reactions are the dominant source of
the stellar luminosity and the main driver of evolution during most
of the core hydrogen-burning phase, the CNO cycles play an
increasingly important role as evolution progresses beyond the
main sequence phase. 
This is because, as temperatures increase, carbon is produced
by the highly temperature-sensitive triple-$\alpha$ reaction and
because, at high temperatures, only a small abundance of CNO
elements is needed for CNO cycle reactions to become the dominant
driver of evolution.
This characteristic behavior of the evolution of zero metallicity
stars has been explored by many authors, from the early works of
\citet{wag74}, \citet{dan82} and \citet{gue83} through more recent
works restricted to the low and intermediate mass stars of concern
here \citep{wei00,mar01,chi01,sch01,sch02,sie02}.  
     In particular, \citet{fuj90}, \citet{hol90} and \citet{fuj00}
have shown that low and intermediate mass stars follow
evolutionary trajectories significantly different from those
followed by population I and population II stars,
becoming carbon stars at a much earlier stage. 

Since the early 1990's, the nature of the evolution of metal poor
stars has been illuminated considerably by the observations.  
Studies of the abundance patterns found in stellar spectra have
revealed that carbon stars are not rare among metal poor stars
and that the relative frequency of such stars increases with
decreasing metallicity, amounting to more than $\sim 20 \%$
for stars with $\feoh \lesssim -2.5$ \citep{ros98}.  
It has been known for a long time that most population I carbon
stars which are not evolved beyond the first red giant branch are
in binaries with a white dwarf companion \citep{mcc80};  this
is also true of population II CH stars \citep{mcc84}.  
     Thus, the frequency of carbon stars among low metallicity
stars may suggest a high rate of binary star formation in
the gas clouds out of which such stars were formed.  
A current lack of evidence for radial velocity variations does not
necessarily rule out the possibility that the observed star is now
in a very wide binary or that the initial binary was disrupted at
some point after the mass-transfer episode occurred.  
     Since an initial binary must be wide enough to accommodate
an AGB star which does not overflow its Roche lobe, and since,
if anything, the binary becomes wider as the AGB star loses mass
\citep[see, e.g., the discussion in][]{ibe00}, it is probable
that the binary separation does not decrease. 
     Furthermore, given the fact that wide binaries are relatively
easily disrupted in close stellar encounters \citep{heg75} ,
it may happen that, in $10^{10}$ yr of traveling through the Galaxy, \mmps\ has lost a one-time companion.
     The radial velocity of \mmps\ has recently been determined
to be $44.5 \pm 0.5$ km s$^{-1}$, based on observations over a 373 
day time span \citep{bes04}.
Although significant radial velocity variations have not been detected,
observations over a much longer baseline would be well worth undertaking.

Extremely metal-poor stars show a large dispersion in the
abundance ratios among metals. The fact that, for
metallicities less than $\feoh \simeq -2.5$, this dispersion
increases with decreasing
metallicity may be interpreted to mean that these stars
were born from interstellar gas which was polluted by a
small number of supernovae, so that stellar abundances reflect
intrinsic variations in the yields of those individual supernovae
that contribute to the pollution of the gas.   

The abundances of many neutron-rich elements vary considerably
from star to star even among those stars which have the abundance
ratio $[\nucm{}{Ba}/ \nucm{}{Eu}]$ very close to the solar
r-process value \citep{mcw98,aok02,hon04}.  
     This fact has been used to argue that the Ba
found in the spectra of stars of low metallicity has been created
by an r-process mechanism in supernovae and that the production
rate varies considerably from one supernova to another \citep{mcw98}.  
     However, there are also large variations from one star to
another in the abundances of those neutron-rich elements which are
unquestionably made by the s-process \citep[see, e.g.,]
[and references therein]{aok01,rya03}.  
     It is known that thermally pulsing AGB stars
are the most likely sites for the formation of heavy s-process
elements.  
     Since most extremely metal-poor stars are not AGB stars,
it seems plausible to invoke the erstwhile presence of an initially
more massive binary
companion which has produced the s-process elements, dredged these
elements to the surface, and then transferred enriched matter to
the less evolved component, a scenario suggested by
\citet{fuj00} and explored by \cite{gor01}, \cite{aok01},
and \citet{iwa04}. These authors point out that s-process
nucleosynthesis in very metal-poor stars may differ significantly
from that which is thought to operate in younger populations.  

The purpose of the present paper is to review the possible scenarios
for explaining the observed properties of \mmps\ and to investigate
the possibility of discriminating among these scenarios on the
basis of our current understanding of nucleosynthesis in
extremely metal-poor stars --- an understanding which has been
achieved by comparing results of theoretical evolutionary
calculations with the observations.  
     In the following, we define stars of metallicity less
than $\feoh \lesssim -2.5$ as ``extremely metal poor'' or
EMP stars and examine the implications of assuming that \mmps\
was born as a single star or as a low-mass component in a
binary system.  
     We use the results of new evolutionary models of Pop.~III
stars of mass in the range $0.8 - 4 \msun$ to elaborate the
binary scenario more fully.  
     Stellar model construction employs up-to-date input physics
\citep{sud03} which differs in significant ways from the
input physics employed by \citet{fuj90} and by \citet{fuj00},
and the results, correspondingly, differ quantitatively in
significant ways.
     In addition, we rely upon new results concerning
neutron-capture nucleosynthesis in a helium-flash driven
convection zone that is a consequence of the ingestion of protons
into this convection zone (Aikawa, Nishimura, Suda,
Fujimoto \& Iben 2004, in preparation).     

In discussing the origin of \mmps, it is important to consider
separately (a) the source of iron group elements, (b) the source
of light elements such as the $\alpha$-rich elements C and O and
the secondary elements N and Na, and (c) the source of s-process
elements.  
     The two possibilities for explaining the metals are: 
     (1) accretion of metal-rich gas by a first
generation (Pop.~III) star; 
     and (2) formation as a second generation star out of
matter in which metals are present in consequence of
mixing with the ejectum of a first generation supernova. 
     In the binary scenario, the enhancements of CNO
elements and of s-process elements are assumed to be due to
accretion from an AGB companion which has experienced one or
more dredge-up episodes that bring to the surface the result
of nucleosynthesis and mixing in its interior. 

There is not yet a consensus as to the conditions which are
necessary for the formation of low metallicity, low mass stars.  
     The lack of a sufficient abundance of metals to provide
a straightforward and effective cooling mechanism for metallicities
$\feoh \lesssim -4$ \citep[]
[see also Omukai 2000]{yos80} is the primary hurdle facing
low mass star formation in the early Universe. 
In the absence of metals, the hydrogen molecule can produce cooling,
but estimates of the minimum Jean's mass vary from $\sim 1 \msun$
\citep{sab77} and $\sim 0.1 \msun$ \citep{pal83} to
$\gtrsim 60 \msun$ \citep{yon72}.
Several authors argue that low
mass stars can form from a metal-free primordial gas cloud \citep[e.g.,][see also Uehara et al.~1996]{ree76},
whereas others suggest that
only massive stars can form from such a cloud \citep[e.g.,][]{omu98,bro99}.
To complicate the picture even further, two dimensional hydrodynamical
simulations \citep{nak01,nak02} suggest a bimodal initial mass
function for Pop.~III stars.  
     Such complexity may arise from the fact that, in primordial
clouds of mass $\sim 10^6 \msun$ which first collapse
\citep[e.g.,][]{teg97}, the scarcity of free electrons
\citep[mole density $\sim 10^{-4} $;][]{gal98} limits the
formation of ${\rm H}_2$ molecules which can act as cooling agents.  
     On the other hand, if the cloud temperature is raised above
$10^4$K and gas is re-ionized, the ionization fraction remains
quite high during the subsequent radiative cooling phase and
increases the abundances of ${\rm H}_2$ and HD sufficiently to
allow the birth of low mass stars even in the complete absence of
metals (Shapiro \& Kang 1987; see Uehara \& Inutsuka 2000
for a recent computation including HD molecules).
     This may be the case for primordial clouds as massive as
$ \gtrsim 10^8 \msun$ for which virial temperatures are
larger than $10^4$ K, although the collapse is delayed somewhat
compared with the case of less massive clouds \citep[see e.g.,][]{nis99}.  
The same situation also prevails even for less massive
primordial clouds when the interstellar gas is swept up and heated
by the supernova shock produced by a first generation massive
star \citep{mac04}. 
     It is also argued that low mass star formation is possible
in clouds of the primordial composition if matter is irradiated by
sufficiently strong FUV radiation from first generation massive
stars \citep{hai96b,omu03}.   
     If the observed enrichment of CNO elements can be explained as acquired after birth, the very existence of \mmps\ indicates that some of these or other formation mechanisms of low mass stars have to work even in the gas clouds completely devoid of metals.   

Observationally, there has been long, continuous interest in search for stars of lower metallicity and/or completely devoid of metals since the first discovery of stars of $1/10 \sim 1/100$ solar metallicity in the 1950's \citep{cha51}. 
     At the beginning of the 1970's, a survey up to a limiting
magnitude of $B \simeq 11.5$ found no star with $\feoh < -3$
\citep{bon70}, and encouraged the idea that no star of
metallicity smaller than this had been formed in our Universe.
     In the early 1980's, however, as by-products of the study
of high-velocity stars and blue subdwarfs, the dwarf G64-12 with
$\feoh = -3.5$ \citep{car81} and the giant CD-38$^{\circ}$245
with $\feoh = -4.5$ \citep{bes84} were found, at magnitudes,
respectively, of $B = 11.8$ and $B = 12.0$, both below the
limiting magnitude of the survey by \citet{bon70}.
     In the early 1990's, the HK survey \citep{bee92}, with a
limiting magnitude of $B = 15.5$, uncovered over 100 stars with
$\feoh < -3$, and, yet, it found no star with $\feoh < -4.0$.  
     As an aside, we note that the metallicity of CD-38$^{\circ}$245
has been revised upwards to $\feoh = - 3.92$ \citep{rya96}. 
     Some time ago, a high velocity carbon dwarf G77-61,
at $B = 15.60$ ($V = 13.90$), was assigned a metallicity of
$\feoh = -5.6$ \citep{gas88}; however, the low surface temperature
of the star and the presence of crowded molecular lines makes
the abundance analysis very difficult, and further work is required
to confirm the early estimate of metallicity. 
     Finally, the Hamburg/ESO survey, with a limiting magnitude
of $B = 17.5$ \citep{chr99}, discovered \mmps\ with metallicity
$\feoh=-5.3$ at magnitude $B = 15.89$. 
Along with the increase in the limiting magnitude, therefore, we have been able to detect the stars of smaller metallicity.
     Provided that the sample of stars of metallicity as small as, and still smaller than, that
of \mmps\ can be considerably augmented, the discovery of \mmps\
may be the beginning of a new epoch in the study of the early history
of the Universe using low mass stellar survivors as a tool.

The paper is organized as follows: In \S 2, formation scenarios
are discussed.  
     In \S 3, the characteristics of evolution and nucleosynthesis
in low and intermediate mass metal-free and metal-poor stars are
reviewed and compared with the results of recent observations
of extremely metal-poor stars.  
     In \S 4, the origin of \mmps\ is discussed and the
binary scenario is elaborated. Conclusions and further
discussion are provided in \S 5.  

\section{Possible Formation Scenarios of HE0107-5240}

The origin of metals in HE0107-5240 must be the consequence of
pollution.  
It is reasonable to suppose that the low mass star currently
observed was formed in a primordial cloud composed initially
of matter in the abundance distribution produced by the Big Bang. 
If it was born before, or simultaneously with, the progenitor of
the first supernova which occurred in this cloud,
we refer to it hereinafter as a first generation (Pop.~III) star;
after birth, it may have accreted matter polluted by the ejectum of
the first supernova. 
If it was born after the first supernova occurred in the cloud,
we refer to it hereinafter as a second generation star;
the matter out of which it was formed was already polluted by the
ejectum of a supernova and subsequent accretion from the polluted
cloud would not significantly alter the surface abundances of the
star.  In either case, while traveling for $10^{10}$ yr through
our Galaxy, the currently observed star must have continued to
accrete matter from the Galactic interstellar medium. Thus,
there are at least three distinct possibilities for explaining
the observed metallicity:
(1) a first generation star which accretes polluted material
in a primordial cloud; (2) a second generation star formed in a
primordial cloud; and (3) a first or second generation star which
acquires most of its metals by accretion from the Galactic
interstellar medium.

We examine first the (third) possibility that accretion from
interstellar gas clouds composed of matter with abundances in
Pop.~I and Pop.~II distributions could by itself have produced
the surface abundances in \mmps. For an order of magnitude
estimate of the accretion rate $\dot{M}$, we adopt the
gravitational focusing cross section
\citep{hoy39}
\begin{equation}
      \sigma_{\rm acc} =   \pi   b_{\rm crit}^2, 
\label{gfcrosssection}
\end{equation}
where the critical impact parameter is given by
\begin{equation}
      b_{\rm crit} = {2 G M\over v^2}. 
\label{impact}
\end{equation}
In this equation, $G$ is the gravitational constant, $M$ is the
mass of the star, and $v$ is the velocity of the star relative
to the velocity of the gas cloud through which it is passing.
Calling $n$ the average number density in a cloud, assumed to
be of uniform density and temperature throughout, and calling
$\mu\ m_{\rm a}$ the average mass of a cloud particle, where
$m_{\rm a}$ is the atomic mass unit and $\mu$ is a number
of the order of unity, the mass accretion rate is
\begin{eqnarray}
  \dot{M}_{\rm acc} & = & n \ \mu\ m_{\rm a}\ \sigma_{\rm acc}\ v \\ 
& = & 5.9 \times 10^{-12}\ \mu\ \left(n \over {\rm cm}^{-3}\right)\
           \left({M \over \msun} \right)^2\
 \left({\hbox{km s}^{-1}\over v}\right)^{3} \ \msun \hbox{ yr}^{-1}.
\label{bondi}
\end{eqnarray}
\citet{bon44} arrive at a value smaller than this by a factor
of $2.5/4$. \citet{bon52} replaces $v$ in eq.~(\ref{impact}) by
$\sqrt{v^2+c_{\rm s}^2}$, where $c_{\rm s}$ is the average speed
of sound in a cloud, basically to avoid a formal divergence when
$v$ vanishes. Given that no version of the simple accretion
formulae is other than a crude approximation, and given the fact
that we can make only order of magnitude estimates of $n$ and $v$,
we settle on
\begin{equation}
      \dot{M}_{\rm acc} = 4 \times 10^{-18}\ n \ M_*^2 \
     v_{100}^{-3} \ \msun \hbox{ yr}^{-1},
\label{bondiacc}
\end{equation}
where $n$ is the particle density in cm$^{-3}$,
$M_*$ is the mass of the star in solar units, and $v_{100}$ is the
relative velocity in units of $100\ \hbox{\rm km\ s}^{-1}$.

The typical space velocity of a low metallicity star is now
of the order of $100 \hbox{ km s}^{-1}$, and this should not
have changed appreciably over the last $10^{10}$ yr. 
If, in the primordial Galaxy, $10^{11} M_\odot$ of baryonic
matter were distributed uniformly in a sphere of radius 10 kpc,
the particle density would be of the order of
$n \sim 1\ \hbox{\rm cm}^{-3}$.
If the gas in the disk of the current Galaxy is of mass
$\sim 10^9 M_\odot$ and if the disk is of thickness $200$ pc
and radius 10 kpc, the mean density is also $\sim 1 \hbox{ cm}^{-3}$.  
We therefore assume that, over a time $T$ and path length
$l=\int_0^T{ d l }$, the average density of matter encountered
by a star is $\bar{n} = \int_0^T{n d l} / l \sim 1 \hbox{ cm}^{-3}$. 
From eq.~(\ref{bondiacc}), we estimate that, over a time
$T = 10^{10}$ yr, an $0.8 \msun$ star accretes a mass of about
$2.6 \times 10^{-8} \msun$.  
Generously assuming that the metallicity of matter in the average
cloud encountered is solar, and adopting $0.2 \msun$ as the mass
of the convective envelope in a metal-poor red giant, we obtain
$\feoh \simeq - 6.7 + 2\ \log M_*$ for the surface metallicity
after the primordial star has become a red giant, far too small
to account for \mmps.  
Adopting $\feoh \simeq -1$ as a more reasonable choice for the
average metallicity of accreted matter, an average particle
density larger than $\sim 200 / M_*^2 \hbox{ cm}^{-3}$ is required
to reproduce the metallicity of \mmps. 
Thus, accretion from interstellar matter during oscillations back
and forth in the Galaxy does not appear as a very promising
explanation for the metallicity of any metal-poor star, let alone
\mmps.

We examine next the first possibility: birth of a first generation
low mass star (possibly in a binary) in a primordial cloud before
it has been polluted, and subsequent accretion of matter in the
cloud which has been mixed with the ejecta of supernovae. 
In the current bottom-up scenarios of structure formation such
as cold dark matter models, the first collapsed objects should have
mass scales of $\sim 10^6 \msun$ \citep[e.g.,][]{hai96,teg97},
in which baryons contribute $\sim 10\%$ of the matter. 
If we assume $v \approx 10 \hbox{ km s}^{-1}$ and
$n \approx 10 \hbox{ cm}^{-3}$, eq.~(\ref{bondiacc}) gives
\begin{equation}
\dot{M}_{\rm acc} = 4 \times 10^{-14}\ n_{10} \ M_*^2 \ v_{10}^{-3}
   \ \msun \hbox{ yr}^{-1}, 
\end{equation}
where $n_{10}$ is the number density in units of
$10 \hbox{ cm}^{-3}$ and $v_{10}$ is the relative velocity in units
of $10 \hbox{ km s}^{-1}$.  
     When mixed with the ejectum of a first supernova
(assuming the mass yield of iron group elements to be
$M_{\rm Fe} = 0.1 \sim 1\ \msun$), the metallicity of the
first collapsed cloud becomes $\feoh = -3 \sim -2$; 
this, of course, increases as more supernovae occur.  
     If the accreting star remains in the cloud for $T \sim 10^9$ yr,
it accretes a mass of $M_{\rm acc} \sim {10^{-4} \msun}$, which is
several thousand times larger than the mass it is expected to accrete
from the Galactic interstellar medium after the primordial cloud has
mixed (dissolved) into the Galaxy.  
    If the time averaged metallicity of the cloud is $\feoh \simeq -2$,
then, when the accreting star becomes a red giant with a convective
envelope of mass $\sim 0.2 M_\odot$, the surface metallicity of the
star reproduces that of \mmps. 

We examine finally the second possibility: a second generation star
born from gas which has been polluted by the mixing of primordial
matter with the ejectum of a first generation supernova. 
As the shock wave produced by a supernova sweeps through the cloud,
gas is compressed into a shell in which possibly low mass stars
can form \citep{mac04}. 
Estimating $\sim 10^5 \msun$ as the mass of intra-cloud gas which
is swept up in the primordial cloud and adopting a nucleosynthetic
yield of $M_{\rm Fe} = 0.1 \sim 1 \msun$ for type II supernovae,
an average value of $\feoh = -3 \sim -2$ can be estimated for
second generation stars.
In this scenario, the large dispersion in the surface abundance
ratios of metals among metal-poor stars of metallicity
around $\feoh \lesssim -2.5$,
found by the HK survey \citep{bee92}, may be attributed to a
dispersion in the abundance distribution produced by individual
supernovae. 
However, in this standard scenario, stars with metallicity as
small as that of \mmps\ are not likely to be formed.  
     Accordingly, only if the supernona ejectum is diluted
somehow in a primordial condensation more massive than
$\sim 10^7 \msun$ (in baryonic mass), can the desired
metallicity be achieved.  
     It is true that low-luminosity supernovae are observed with
an Fe yield $\sim 100$ times less than that of a typical supernova,
but it is hard to trigger star formation since, because of low
energy, the supernova shock is dissolved before initiating the
fragmentation in the swept-up gas shell \citep{mac04}.  

In closing this section, we comment on the scenario for the
formation of \mmps\ proposed by \citet{ume03}. 
They show that, except for the abundance of nitrogen and sodium,
the observed element abundances of HE0107-5240 can be explained
by assuming that the star was born out of matter produced by mixing
the ejectum of an unusual Type II supernova with a judiciously
chosen mass of primordial matter.
They argue that the nitrogen not produced by the supernova model
appears at the surface of the present star in consequence of a
first dredge-up episode on the red giant branch.  
    As will be discussed in \S4, we have calculated the
evolution of a model star of mass $0.8\msun$ and of a chemical
composition suggested by \cite{ume03} and find that, indeed,
the $[\nucm{}{N}/\nucm{}{Fe}]$ ratio in \mmps \ can be reproduced
during the first dredge-up episode.
     Thus, one important aspect of the Umeda-Nomoto scenario
is confirmed.  
     However, the enrichment of Na has yet to be explained.  
     In addition, when we consider the likelihood of the formation
of a low mass star from the polluted cloud postulated by
\citet{ume03}, several problems arise.
     The mass of carbon ejected by the postulated
supernova is $M_{\rm C} \simeq 0.2\ \msun$, implying that the
mass of primordial matter mixed with the ejectum is only
\begin{equation}
     M_{\rm shell} = {M_{\rm  C}\over  X_{\rm C, \odot} 
     \times 10^{[C/H]}} \approx 1.3 \times 10^3 \ \msun,
\end{equation}
where $X_{\rm C, \odot}=3.0 \times 10^{-3}$ is the solar abundance
by mass of carbon and $[{\rm C}/{\rm H}]=-1.3$ is the carbon
abundance in \mmps.
This is a factor of $\sim 100$ smaller than the mass which must
be swept up into a gas shell by a first generation supernova before
fragmentation is expected to occur in the shell \citep{mac04}. 
     Thus, star formation in the swept-up gas shell is unlikely.
     If, despite these arguments, star formation can occur within
such a small shell, one might anticipate that pollution by
normal supernovae with similar or larger mass yields of iron
group elements would give rise to the formation of stars of
metallicity $\feoh \gtrsim -1$.  
     This, however, leads to difficulty in explaining the origin
of extremely metal-poor stars of metallicity $\feoh \simeq -3$.

\medskip

In summary, the viable scenarios for the formation of \mmps\ are:

\noindent (1) A first generation star which \hfill\break
\indent (a) disguises its surface by
accreting metals in the parent primordial cloud polluted by the
ejectum of a normal supernova and then \hfill\break
\indent (b) accretes, from an evolved
companion in a binary system, matter containing CNO and s-process
elements produced in the interior of the companion.

\noindent (2) A second generation star which
\hfill\break 
\indent (a) forms in a primordial cloud of total
(dark matter plus baryonic matter) mass $\gtrsim 10^8 \msun$
polluted by the ejectum of a normal supernova,
and acquires CNO elements from an evolved binary companion, or
\hfill\break
\indent (b) forms from material polluted by the ejectum of a very
unusual supernova which makes C, O, and Na in the currently
observed ratios. 

\smallskip

In the following, we examine how one may choose from among these
scenarios by using knowledge gained from a study of the evolution
of metal-free and metal-poor model stars.   

\section{Evolutionary Characteristics of Low and Intermediate Mass
Metal-Poor Stars}
Many numerical computations of the evolution of low and intermediate
mass Pop.~III and EMP stars and discussions of their
characteristics have been published 
\citep{wag74,dan82,gue83,fuj90,hol90,cas93,fuj95,cas96,fuj00,wei00,mar01,chi01,sch01,sch02,sie02}.  
It is commonly accepted that, as first shown by \citet{fuj90},
a carbon enhancement occurs at the core helium flash stage in
low mass model stars.   
Evolutionary destinations are illustrated by Fig.~2 of \citet[]
[hereinafter FII00]{fuj00} as a function of initial metallicity
and mass.  
Though some differences in the evolution of low-mass Pop.~III stars
are reported by other groups \citep{sch02,sie02}, qualitative
results are not different from those found by FII00. 

Nevertheless, because there are many important aspects of
nucleosynthesis that have not yet been adequately addressed,
we have computed additional models of zero metallicity stars
of mass ranging from 0.8 to 4.0 $\msun$.   
     We have used the stellar evolution code employed by FII00
\citep[see e.g.][]{ibe92}, modified with up-to-date input
physics \citep{sud03};  
     radiative opacities are taken from OPAL tables \citep{igl96}
and \citet{ale94} tables, complemented in regions not covered by
the tables by analytic approximations \citep{ibe75}. \citet{ito83}
is adopted for electron conductivity in dense regions of the
stellar interior;   
     elsewhere, conductive opacities are calculated with a
fitting formula \citep{ibe75} based on \citet{hub69} \citep
[for details of interpolation, see][]{sud03}.  
     Neutrino loss rates are taken from fitting formulae by
\citet{ito96}.  
     The NACRE compilation \citep{ang99} is used for nuclear
reaction rates involving proton, alpha, and electron capture
for \nuc{1}{H}, \nuc{3}{He}, \nuc{4}{He}, \nuc{12}{C}, \nuc{14}{N},
\nuc{16}{O}, \nuc{18}{O}, \nuc{22}{Ne}, and \nuc{25}{Mg}.  
     In treating convection, we adopt the standard mixing-length
formulation, choosing the parameter $\alpha = 1.5$, and use the
Schwarzschild criterion for instability against convection.  

For models of initial mass $M \leq 1.1 \msun$, the star reaches
the red giant branch (RGB) with an electron-degenerate helium
core before helium ignites off-center in the core and forces the
development of a convective shell.
    The entropy barrier between the helium convective region
and the hydrogen-rich envelope is small. Thus, as it extends outward,
the outer edge of the convective region generated by the off-center
helium core flash penetrates the hydrogen profile. Hydrogen is
mixed into the convective shell and a hydrogen-shell flash is ignited.
     The initial convective region splits into two convective zones,
the inner one driven by the helium core flash and the outer one
by the hydrogen shell flash.   
     The convection zone sustained by hydrogen burning extends
outward, adding fuel by ingesting protons.  
     These developments are qualitatively consistent with FII00,
though quantitative results are different.  
     The expansion of the hydrogen-flash driven convection zone
brings about the movement inward in mass of the base of the
surface convective zone and the dredge-up of nuclear matter
processed in the hydrogen mixed region, changing the surface
chemical composition dramatically.  
     Hereinafter, we refer to these processes, triggered by
hydrogen mixing into a helium-flash driven convective zone, as
He-FDDM (Helium-Flash Driven Deep Mixing).

For $M \geq 1.2 \msun$, helium ignites at the center before
electron degeneracy has set in, and the model avoids the RGB
configuration (stable hydrogen-burning in a shell above an
electron-degenerate helium core). After helium is exhausted
at the center, the star adopts an early asymptotic giant branch
(EAGB) configuration consisting of an electron-degenerate
carbon-oxygen (CO) core capped by a helium-rich shell (at the
base of which helium is initially burning) and a hydrogen-rich
envelope (at the base of which hydrogen is not initially burning).
Almost immediately the models evolve into thermally pulsing AGB
(TPAGB) stars with nuclear energy being produced alternately
by helium burning and hydrogen burning. For models of mass
$1.2 \msun \leq M \leq 3 \msun$, a He-FDDM episode initiates
the TPAGB phase when the first helium convective region touches
the hydrogen profile. The subsequent ingestion of hydrogen into
the helium convective zone and the formation of two convective zones
followed by dredge-up of nuclearly processed material into the
envelope is expected to proceed just as in the case of low mass
models which ignite helium under electron-degenerate conditions.
The He-FDDM episode does not occur in the $M = 4 \msun$ model.

The duration of the He-FDDM event is so short that our
treatment of convective mixing is not correct, so, in both the
RGB and AGB cases, we have terminated our computations at the
start of this event.  
     A correct time-dependent study of He-FDDM has been
conducted by \citet{hol90}, and we assume that the qualitative
behavior of all such events is essentially invariant.
Details of the evolution after the mixing episode will
be provided in a subsequent paper (Suda, Fujimoto \& Iben, in
preparation).  

Figure~1 shows a selection of evolutionary tracks of Pop.~III
models in the H-R diagram. 
     For stars of $M \le 3.0 \msun$, evolution is terminated at
the point when models begin the He-FDDM episode which turns them
into carbon stars.  
     Compared with Pop.~I and Pop.~II model stars, Pop.~III model
stars become carbon stars at significantly smaller luminosities,
and hence, smaller radii.    

Recently, \citet{chi01} and \citet{sie02} have computed the evolution
of Pop.~III stars of mass $M \ge 4 \msun$, forcing mixing
across the boundary between the carbon-rich helium layer and
the hydrogen-rich layer which is formed when the outer edge of the
convective zone engendered by helium burning approaches the hydrogen-rich
layer.  
     They contend that this ``carbon injection'' mechanism is
different from the He-FDDM mechanism and is caused by
convective overshoot.
     Because of a huge jump in the carbon abundance across the
boundary, by many orders of magnitude, the mixing of carbon into
the hydrogen-rich layer enormously enhances the hydrogen-burning rate,
giving rise to an ``irreversible'' process \citep{sie02}.
     The initial increase in the surface carbon abundance leads to
full amplitude thermal pulses and third dredge-up episodes similar
to those in Pop.~I and II stars, resulting in surface enhancements of
C, N, and O similar to those observed in \mmps; the oxygen enhancement
is due to the reaction $\nucm{12}{C} (\alpha, \gamma) \nucm{16}{O}$
at high temperatures (see the discussion in \S3.3).   
     These results, however, may depend critically on the adopted
assumptions, and in particular, may be an artifact of the adopted
prescription for convective overshoot.   

We compute the evolution of a $4 \msun$ star without any special
treatment of convective overshooting.  
     In addition, we treat changes in chemical composition
explicitly, with small time steps, so as to maintain consistency
with the associated energy-generation rates; the difference in
the changes in chemical composition may affect the temperature
structure in the helium core through compression.  
     Figure~2 shows the resultant variations in the helium-burning
luminosity.
     We confirm that thermal pulses do occur, in agreement with
\citet{sie02}, but in disagreement with results of earlier
computations \citep{chi84} and of a related
semi-analytical study \citep{fuj84}.  
These discrepancies must be due to the sensitivity of results
to the input physics.

The TPAGB evolution of our $4 \msun$ model star
starts when the core mass is $M_{\rm c} = 0.873 \msun$, and we
have followed 23 thermal pulses until a locally asymptotic
strength \citep{fuj79} of $L_{\rm He}^{\rm max} =
1.9 \times 10^5 L_\odot$ is reached.  
      In contrast with the Seiss et al. calculations, we do not
find any direct mixing between the carbon-rich and hydrogen-rich
layer.  
     The minimum distance between the outer edge of the helium
flash driven convective zone and the base of the hydrogen profile
is always larger than a pressure scale height ($\sim 1.4 H_P$).  

Further investigations are necessary to determine the relevance of
intermediate mass metal-poor models to HE0107-5240, as well as to
determine why, in contrast with such models constructed two decades ago,
current models experience thermal pulses.

\subsection{C and N Enhancements}

As described in the evolutionary diagram in FII00, and consistent
with the results we have obtained here, the evolution of
low and intermediate mass stars may be classified into the
following groups.  
     For $M \leq 1.1 \msun$ and $\feoh \lesssim -4.5$
(Case I, FII00), the model becomes an RGB star before experiencing
the He-FDDM mechanism;
     For $M \leq 1.1 \msun$ and
 $-4.5 \lesssim \feoh \lesssim -2.5$ (Case II), a model experiences
the He-FDDM at the start of the TPAGB phase.  
     If stellar mass is in the range $1.2 \leq M/\msun{} \lesssim 3$
and $\feoh \lesssim -2.5$ (Case II$^{\prime}$), models undergo the
He-FDDM during the EAGB phase.  
     As a consequence of the dredge-up of matter which has been
processed by helium burning and then processed by proton capture
reactions, metal-poor stars become nitrogen-rich carbon stars.  
     In particular, low mass stars of the Case I and Case II varieties
increase their surface carbon and nitrogen abundances to the extent
that $[{\rm C}+ {\rm N} / {\rm H}] = 0 \sim -1$ during a single
He-FDDM mixing episode, with ${\rm C} / {\rm N} \simeq 1 \sim 1/5$
for Case I and Case II, respectively \citep{hol90,fuj00,sch02}.  
     In Case II models, the abundance ratio of C to N in
the hydrogen flash convective zone is larger than in Case I
models since, owing to a larger mass, and hence, a deeper
gravitational potential of the core, more helium has to be burned
(yielding a larger C abundance) before the outer edge of the helium
convective zone reaches the base of the hydrogen-rich layer.  
     In Case II$^{\prime}$ models, on the other hand, the enhancement
of C and N due to the He-FDDM event is relatively small since (1)
the mass of the helium convective zone, and hence, the amount of C-
and N-rich matter dredged up, decreases with increasing core mass,
and (2) the dilution in the envelope increases with increasing
envelope mass.   
     After the surface abundance is increased above
$ [{\rm C N O}/{\rm H}] \simeq -2.5$, the model stars behave
similarly to those with metallicity $\feoh \gtrsim -2.5$. This
is because the structure of the hydrogen-burning shell during
the TPAGB stage is controlled mainly by the hydrogen shell
burning rate which depends only on the CNO abundance;
the opacity in the shell is dominated by electron scattering
and therefore depends very little on the metallicity.  
     Accordingly, the helium-flash convective zone cannot
reach the base of the hydrogen-rich shell, and the
He-FDDM mechanism does not occur. The models enter the
TPAGB evolutionary phase in the same fashion as do model
stars of younger populations (FII00); successive third
dredge-up episodes increase the surface
C abundance and lead to a large C/N abundance ratio.

\subsection{Neutron Capture Nucleosynthesis} 

The character of mixing during a He-FDDM event, as studied
in detail by \citet{hol90}, is illustrated schematically
in Figure 3(a).  
     As the helium shell flash develops, the helium convective zone
grows in mass until, near the peak of the flash, its outer edge
extends into the hydrogen-rich layer.
     Hydrogen is then carried downward by convection until
it reaches a point where the lifetime of a proton becomes less
than the convective mixing timescale; at this point, hydrogen burns
via the $\nucm{12}{C} (p, \gamma) \nucm{13}{N}$ reaction.
     With the additional energy flux produced by this burning
reaction, the shell convective zone splits into two parts, the outer
one driven by the energy flux due to hydrogen burning and the
inner one driven by the energy flux due to helium burning.  
     The convective shell engendered by hydrogen burning transports
C- and N-rich matter outwards where it can ultimately be incorporated
into the envelope convective zone.
     In models with large core masses, the convective shell
sustained by helium burning persists even after the hydrogen-driven
convective zone has disappeared (FII00; Iwamoto et al.~2004).  

Before the split into two parts of the initial, helium-driven
convective shell occurs, some of the mixed-in hydrogen which has
been captured by \nuc{12}C continues inward and, after the split,
is trapped in the surviving helium convective zone as \nuc{13}{N}
and/or as its daughter \nuc{13}C.  
      At the high temperatures in the surviving helium convective shell,
the reaction $\nucm{13}{C} (\alpha, n) \nucm{16}{O}$ occurs, with
interesting consequences, as demonstrated by \citet{iwa04} for a
$2 \msun$ star with $\feoh = -2.7$. 

To explore the ensuing neutron-capture nucleosynthesis in the
helium convective shell, we adopt the same one-zone approximation
used by \citet{aik01}.  
     Our nuclear network includes 59 isotopes of 16 light species
from the neutron and \nuc{1}H through \nuc{31}{P}. Charged particle
reaction rates are taken from \citet{ang99} and \citet{cau88},
and neutron-capture cross sections are taken from \citet{bao00}.
     The model parameters of shell flashes are taken from the
computation of the evolution of a $2.0 \msun$ star (FII00) in
which hydrogen is ingested by a helium convective zone.  
     In this work, we treat the amount of mixed \nuc{13}C as a
parameter since the amount of \nuc{13}C mixed into the inner helium
convective zone varies with the strength of the helium shell flash,
and hence, with the stellar mass, and, also, depends on the treatment
of convection. The mixing is assumed to occur at the peak of the shell
flash.  

Figure~4 illustrates the progress of nucleosynthesis in the
helium convective zone in a Pop.~III model star, unpolluted with
accreted metals, when the amount of \nuc{13}{C} mixed into this
zone is chosen in such a way that
$\nucm{13}{C}/\nucm{12}{C} = 10^{-3}$
(Aikawa et al.\ 2004 in preparation).  
     As soon as $\nucm{13}{C}$ is mixed into the convective zone, it
rapidly reacts with helium to produce neutrons.   
     The neutrons so produced are captured primarily by $\nucm{12}{C}$
to form $\nucm{13}{C}$ and then reappear in consequence of additional
$\nucm{13}C (\alpha, n) \nucm{16}{O}$ reactions. This neutron
cycling process continues until the $\nucm{16}O (n,\gamma)
\nucm{17}O$ reaction has converted most of the initially injected
$\nucm{13}C$ into \nuc{17}O.  
     Then, $\alpha$ capture on \nuc{17}{O}, which occurs $\sim 10^{4}$
times more slowly than capture on \nuc{13}{C} at the relevant
temperatures, starts to produce neutrons via the reaction
$\nucm{17}O (\alpha, n) \nucm{20}{Ne}$.
     The newly formed \nuc{20}{Ne} consumes neutrons
to yield heavier isotopes of Ne and isotopes of Na and Mg.

The nature of s-process nucleosynthesis depends on whether or not
seed nuclei are present in the helium convective zone.
     In second-generation stars, iron-group elements
act as seeds for the production of heavy s-process isotopes
by neutron capture.
     In first-generation stars, if no iron-group seed
nuclei have entered the helium zone in consequence of the inward
mixing of accreted iron-rich matter, most neutrons are
absorbed by isotopes of Ne, Na, and Mg which are made by successive
neutron captures and beta decays beginning with $^{20}$Ne as a seed
nucleus. The possible presence of iron-group seed nuclei due to
pollution and mixing are discussed in \S 4. 
     The neutron-rich isotopes synthesized in a helium
convective zone remain in place after convection ceases
and some of them are incorporated into subsequent helium
convective zones which appear during thermal pulses, as
illustrated in Figure~3(a).  
     In Case II$^{\prime}$ models, after [C/H] $> -2.5$,
neutron-rich isotopes made in the He-FDDM convective zone are
dredged up to the surface along with carbon during third dredge-up
episodes (FII00, and also see Iwamoto et al.~2004).   

This mechanism for s-process nucleosynthesis works only for
$\feoh \lesssim -2.5$ (FII00), and contrasts with the
mechanism of radiative \nuc{13}{C} burning which may operate
in younger populations, as illustrated in Figure~2b
\citep[Straniero et al.~1995, and see also]
[and references therein]{bus99}.  

The radiative \nuc{13}{C}-burning model invokes, by some
extra-mixing mechanism which is assumed to operate during the
third dredge-up phase, a partial mixing of hydrogen-rich matter
with carbon-rich matter which was formed during the preceding 
helium shell flash in the convective zone driven by helium burning.
In low mass TPAGB stars, the temperatures across the interface
between hydrogen-rich and carbon-rich zones become small enough
during the dredge-up phase that carbon nuclei partially recombine
with electrons, raising the opacity \citep{sac80}; the opacity
becomes proportional to the abundance of $^{12}$C and
semiconvection spreads out the $^{12}$C and $^1$H abundance
profiles so that they overlap, producing in the hydrogen profile
a long tail which descends inwards with a gradual slope
into the carbon-rich region \citep{ibe82a,ibe82b,hol89};  
     finally, when hydrogen re-ignites in this region, proton capture
on $^{12}$C produces a $^{13}$C ``pocket''.
This mechanism works only for low mass stars of low metallicity
\citep{ibe83}; in low mass stars of higher metallicity, the
contribution to the opacity by carbon ions is not large enough
to activate semiconvection; in intermediate mass TPAGB stars,
the temperatures near the carbon-hydrogen interface do not become
small enough for carbon to retain bound electrons. However,
as discussed by \citet{str95},
the observations overwhelmingly suggest
that $\alpha$ capture on $^{13}${C} is the source of neutrons for
s-process nucleosynthesis in low mass Pop.~I stars as well as in
intermediate mass stars of both the Pop.~I and Pop.~II variety,
and that, therefore, some other mixing mechanism must also be operating.  
     It is inferred that some form of convective overshoot
and/or rotation-induced mixing must be capable of producing
a region of overlapping $^{12}$C and $^{1}$H abundances, and
that proton capture on $^{12}$C subsequently produces a
$^{13}$C-rich pocket
in all TPAGB stars (see e.g., Busso et al~1999).
 
Prior to the work of \citet{str95}, it was assumed
that the \nuc{13}{C} in the \nuc{13}{C} pocket survives the
interpulse phase and is incorporated intact into the next
helium flash-driven convective zone, with results similar to
those found in the He-FDDM model
\citep[see e.g.,][]{gal88,hol89,hol90,bus92}.
\citet{str95} find instead that,
due to compression brought about by the increase in core mass,
temperatures in the \nuc{13}C-rich pocket rise sufficiently
during the interpulse phase that \nuc{13}{C} captures alpha
particles and releases neutrons, driving s-process nucleosynthesis
in the pocket, which remains in a radiative zone during the
interpulse phase. The resultant
s-process isotopes are mixed intact into the convective zone
formed during the next helium shell flash, and no further s-process
nucleosynthesis occurs in the convective shell unless, as in
the case of Pop.~I and Pop.~II stars, some $^{22}$Ne is incorporated
into the convective shell.

Using the \nuc{13}{C}-pocket scenario adopted by \citet{gor00}
for younger population stars, \citet{gor01} compute the radiative
\nuc{13}{C} burning in a Pop.~III star and argue that heavy elements
such as Pb can be synthesized even if the stars initially lack
iron-group elements.
     The differences between our results and theirs may stem from
the following points: 
\hfill\break
     (1) We deal with neutron irradiation only once during the
He-FDDM event, while they assume multiple irradiations during the
occurrence of many (22) shell flashes, taken from a $3 \msun$ model
by \citet{sie02}, which, differently from our model, undergoes
carbon injection outward rather than hydrogen mixing inward. 
\hfill\break
     (2) In our convective \nuc{13}{C}-burning model, most neutrons
are first converted into \nuc{17}{O}, and subsequent neutron-liberating
reactions produce Ne isotopes with large neutron-capture cross sections
at an abundance even larger than the abundance of \nuc{13}{C} added,
while in the radiative \nuc{13}{C}-burning model, temperatures
are too small for the reaction $\nucm{17}{O} (\alpha, n)
\nucm{20}{Ne}$ to operate \citep{mat92}.  
     In our models, the Ne isotopes act as effective neutron absorbers,
preventing the formation of heavy s-process elements.
\hfill\break
     (3) The amount of \nuc{13}{C} injected into our convective
zone is significantly smaller than the amount of \nuc{13}{C}
postulated to be in the \nuc{13}{C} pocket of the radiative burning
model. In our model, the separation of convective zones sets a limit
and we assume a relatively small ratio of
$\nucm{13}{C}/\nucm{12}{C} \simeq 10^{-3}$, whereas \citet{gor01}
argue that the main contribution to heavy s-process elements
comes from layers where $10^{-2} \lesssim \nucm{13}{C}/\nucm{12}{C}
\lesssim 1$.  

In any case, the outcome of radiative \nuc{13}{C} burning
depends entirely on the abundance of \nuc{13}{C} in the
postulated \nuc{13}{C}-rich pocket.
     Surely, the properties of the $^{13}$C-rich pocket vary
with the metallicity since, during a third dredge-up episode,
metallicity differences affect the structure at the interface
between the surface convective zone and the carbon-rich layer
left behind by the helium-flash driven convective shell.
     We return to this point in \S 3.4 where we discuss
comparisons with the observations.

\subsection{Oxygen and Sodium Enrichment}

The convective \nuc{13}{C} burning which occurs during the He-FDDM
episode can be an effective mechanism for producing oxygen and sodium
in stars of mass and metallicity in the Case II$^{\prime}$ category.   
     In the absence of other competing neutron absorbers, the
neutron recycling reactions $\nucm{12}{C} (n, \gamma) \nucm{13}{C}
(\alpha, n) \nucm{16}{O}$ \citep{gal88} work effectively. 
     When the \citet{bao00} neutron-capture cross sections
are adopted, the final oxygen abundance produced through these
reactions varies from $4.0 \times 10^{-3}$ to 0.022 as the
abundance ratio of mixed-in \nuc{13}{C} to \nuc{12}{C}
is varied from $\nucm{13}{C}/\nucm{12}{C} = 10^{-4}$ to $10^{-2}$.   
     Correspondingly, the mass ratio of oxygen to carbon
decreases from $\nucm{12}{C}/\nucm{16}{O} \simeq 10^2$ to 
$\nucm{12}{C}/\nucm{16}{O} \sim 10$.   
     The final number of \nuc{16}{O} nuclei produced is
much larger than the number of added \nuc{13}{C} nuclei with, 
on average, each added neutron being captured between $\sim 200$
and $\sim 10$ times by \nuc{12}{C}, depending on the amount
of injected \nuc{13}{C}.

Another consequence of convective \nuc{13}{C} burning is
the enrichment of neon, sodium and magnesium.  
     As in the case of \nuc{16}{O}, the number of \nuc{20}{Ne} 
nuclei produced substantially exceeds the number of mixed-in
\nuc{13}{C} nuclei, since some of the neutrons released via
the $\nucm{17}{O} (\alpha, n) \nucm{20}{Ne}$ reaction are
recaptured by \nuc{12}C and are recycled into
\nuc{17}{O} and thence into \nuc{20}{Ne}.  
     The final Ne abundance in the flash convective zone
varies from $5 \times 10^{-5}$ to $\sim 3 \times 10^{-3}$ as
the initial ratio $\nucm{13}{C}/\nucm{12}{C}$ is varied from
$10^{-4}$ to $\sim 10^{-2}$.
     Subsequent neutron captures on \nuc{20}{Ne} yield
heavier isotopes of Ne, and, ultimately, to isotopes of
Na and Mg.
     The sodium abundance increases from $10^{-7}$ to $10^{-5}$,
nearly in proportion to the abundance of the mixed-in \nuc{13}{C}.

If the total abundance of CNO isotopes in the envelope exceeds
$[{\rm C N O}/{\rm H}] \simeq -2.5$, the outer edge of the helium-flash convective
zone cannot reach the base of the hydrogen-rich shell, and the
He-FDDM mechanism no longer functions.  
     Nevertheless, additional mechanisms operate to enrich
oxygen and sodium during subsequent thermal pulses.  

During a helium shell flash, \nuc{16}{O} is produced by
$\alpha$ capture on the \nuc{12}{C} made by the
$3 \alpha$ reaction.  
     Figure~5 shows, for four different temperatures, the ratio
of the net production rate of \nuc{12}{C} to the production
rate of \nuc{16}{O} by these reactions as a function of $X_{\rm C}$,
the abundance by mass of $^{12}$C.
Along the vertical axis, $\dot{\rm O}$ is the rate of the
$^{12}$C$(\alpha,\gamma)^{16}$O reaction and $\dot{\rm C}$ is
the rate of the triple alpha reaction minus the rate of the
$^{12}$C$(\alpha,\gamma)^{16}$O reaction. For a given temperature,
the ratio of production rates decreases rapidly with increasing
$X_{\rm C}$.  
For a given $X_{\rm C}$, the $\dotc / \doto$ ratio at first increases
with increasing temperature and then decreases, with a sharp maximum
occurring at $T \simeq 2.3\times 10^8$ K.
     For values of $X_{\rm C} \simeq 0.15 \sim 0.2$, typical
in helium convective zones during recurrent helium shell flashes,
if the temperature is not close to the maximum, the $\coopr$
production-rate ratio can be smaller than several times ten,
rapidly decreasing with increasing carbon abundance.

The thin solid curve in Figure~5 describes the evolutionary
variation of the abundance ratio ${\rm C} / {\rm O}$ at the
fixed temperature $T = 2.3 \times 10^8$ K, starting with pure helium. 
     For a given carbon abundance, the evolutionary abundance ratio
is larger by a factor of several (typically $\sim 3$) than the
production-rate ratio for the same carbon abundance.
     Accordingly, after an initial decrease achieved in a
He-FDDM episode, it is possible, during the early stages of
the third dredge-up phase in low mass ($\le 3 \msun$) stars,
to sustain relatively small ${\rm C} / {\rm O}$ abundance ratios
(as small as ${\rm C} / {\rm O} \lesssim \hbox{a few tens}$)
in the helium-flash convective zone where
$X_{\rm C} \simeq 0.15 \sim 0.2$.
     For more massive intermediate-mass stars, which begin TPAGB
evolution with larger core masses, the maximum temperatures
in the helium-flash convective zone are even higher
($T_{\rm max} \gtrsim 4 \times 10^{8}$ K).   
     In such stars, the abundance ratio $\nucm{}{C}/\nucm{}{O}$
in the helium convective zone approaches the production rate
$\coopr$ for $X_{\rm C} \simeq 0.2$
and abundance ratios as small as $\nucm{}{C}/\nucm{}{O} \simeq 10$
can be achieved even without benefiting from a precursor He-FDDM
event.

With regard to Ne and Na isotopes, other enrichment channels
also exist that do not hinge exclusively on the operation of
a mechanism for mixing carbon-rich with hydrogen-rich material
during third dredge-up episodes.
     In the helium flash convective zone, \nuc{22}Ne can be
produced by the reactions $\nucm{14}{N}(\alpha, \gamma) \nucm{18}{F}
(e^+ \nu) \nucm{18}{O} (\alpha, \gamma) \nucm{22}{Ne}$. The
\nuc{14}{N} can be produced in the hydrogen shell-flash
driven convective zone which appears during the He-FDDM event.
If the stellar envelope is already CNO-enriched, \nuc{14}{N}
is produced by quiescent hydrogen burning and is added to
the helium core during the interpulse phase.  
     If the core mass is small and the maximum temperature
remains below $T_{\rm max} = 4 \times 10^8$ K, \nuc{22}{Ne}
survives the flash unburned and is added to the envelope by
the following third dredge-up event, contributing to the
surface enrichment of other neon isotopes formed during
a possible precursor He-FDDM event. 

Sodium can be formed during quiescent hydrogen burning
as some neon isotopes are converted into Na through the
Ne-Na cycle reactions.  
     At the temperatures prevalent in the hydrogen-burning
shell during the early TPAGB phase in the stars of concern here,
the Na abundance produced by Ne-Na chain reactions is rather small
(about a few \% of Ne), and yet, thanks to the large enrichment of
Ne isotopes, this can amount to a significant overabundance of Na
relative to the scaled-solar abundance.  
    When formed, Na can survive the helium flash, be dredged up
by surface convection and contribute to a surface enrichment. 
    It is to be noted that \nuc{24}{Mg} is also produced through
the Ne-Na chain reactions in the quiescent hydrogen-burning shell,
although the amount produced depends strongly on the branching
ratio between the $(p, \alpha)$ and $(p, \gamma)$ reactions on
\nuc{23}Na (Angulo et al.~1999; see also Aikawa et al.~2001).   

As a corollary, O and Na enrichments coupled with C and N
enhancements are a symptom of TPAGB evolution in which both
the He-FDDM mechanism and the third dredge-up mechanism have worked.  

\subsection{Comparisons with the Observations}

Recent observations reveal the peculiar aspects of the most
metal-poor stars.  
     One of the major characteristics of these stars is a large
frequency of carbon stars.  
     This frequency is $\sim 25 $\% for
known stars with \feoh $< -3$, a frequency far larger than
the few percent for Pop.~I and II stars \citep{ros98}.
The Hamburg/ESO objective-prism survey confirms that, among
EMP stars, there is a high frequency of stars with strong
carbon-enhancements (N. Christlieb 2004, private communication).

As pointed out by FII00, the large frequency of carbon stars
among EMP stars can be explained by the occurrence of the
He-FDDM mechanism, which enables such stars to become carbon
stars for a wider mass range and at an earlier phase of evolution
than can stars of younger populations.
     The initial stellar masses of stars which can become carbon stars
in this way span the entire mass range below $\sim 3 \msun$ of stars
that can evolve off the main sequence in a Hubble time, including the
low masses for which the third dredge-up cannot work in younger
populations. 
     Thus, the fraction of EMP stars that can end their
nuclear-burning careers as carbon stars is much larger than
the fraction of Pop.~I and Pop.~II stars which can do so.  
     In addition, because of a smaller core mass and smaller
CNO abundances, when an EMP star evolves into a carbon star, 
its radius is much smaller than that of a carbon star belonging
to younger populations; this means that EMP binary systems
which produce carbon stars can have smaller initial orbital
separations than can younger generation binary systems which
produce carbon stars.
    Most currently observed carbon stars owe their spectral
peculiarity to prior mass transfer from an initially more massive
component in a binary system; this fact, coupled with the larger
fraction of EMP binary systems which can accommodate TPAGB stars and
the larger range of masses for progenitors of EMP carbon-rich
stars (compared with their counterparts among younger
generations) may account for the large fraction of carbon
stars among EMP stars.
    Moreover, many EMP carbon stars exhibit large
enhancements of N and very large enhancements of C,
with $[{\rm C + N} / {\rm Fe}] \gtrsim 2$ and 
 $[{\rm C + N} / {\rm H}] = 0 \sim -1$, consistent
 with theoretical predictions.     

Among EMP carbon stars, there is a correlation between
the abundances of carbon and nitrogen and the abundances of
s-process elements, as pointed out by \citet{aok02}.   
     In a diagram similar to theirs, we show in Figure~6
the Ba or Pb enhancement versus the
$\nucm{}{C} ( + \nucm{}{N})$ enhancement for stars with
metallicities in the range $-3.75 \le \feoh < -2.5$.
     Data has been taken from
\citet{rya91,nor97,nor97b,nor01,hil00,hil02,aok01,aok02,aok02b,dep02,joh02,coh03,luc03,sne03}.  
     Filled circles, open circles, and crosses denote, respectively,
the coordinates $[{\rm Ba} / {\rm Fe}]$ versus
$[{\rm C}+ {\rm N} / {\rm Fe}]$, $[{\rm Pb} / {\rm Fe}]$ versus
$[{\rm C}+ {\rm N} / {\rm Fe}]$, and $[{\rm Ba} / {\rm Fe}]$ 
versus $[{\rm C} / {\rm Fe}]$. 
     In the figure, stars are clearly separated into two
branches, one with and one without s-process element enhancements.
The C (+ N) enhancement covers the same range on both branches,
namely, $[{\rm C} (+ {\rm N}) / {\rm Fe}] \simeq 0 \sim 2.5$.   
     On the s-process-enriched branch, the abundance of s-process
elements increases, with increasing C (+ N) abundance, over a
range of $\sim 100$ or more.
     For stars in the chosen metallicity range, there are two
different evolutionary paths to the carbon stars, Case II and
Case II$^\prime$, which are separated by the nature of enhancements
during third dredge-up episodes following the occurrence of a
He-FDDM event.
     These two paths may correspond to the two branches evident
in Figure~6, Case II to the branch without s-process element
enrichment and Case II$^\prime$ to that with s-process element
enrichment. 
     The theoretically predicted abundance characteristics
for each case, indicated by the large open circles in Figure~6,
actually coincide with the abundance characteristics at the
rightmost end of each branch.   
     In Case II models, the C and N enrichments are produced
only by the He-FDDM mechanism, with a relatively large nitrogen
enrichment and without a significant enhancement of s-process
elements.  
     In Case II$^\prime$ models, third dredge-up episodes,
following a He-FDDM episode, give rise to the enhancement
of s-process elements as well as to a decrease in the
${\rm N}/{\rm C}$ ratio because of the production of
\nuc{12}{C} by the $3 \alpha$ reaction.   
     This is in agreement with the fact, pointed out by
\citet{rya03}, that the degree of processing by CN-cycle reactions
is clearly greater in stars without s-process element enhancements
than in stars with s-process element enhancements.   

It has been also pointed out \citep{aok01,aok02b,rya01} that metal
poor stars exhibit interesting variations in the relative
abundances of s-process elements of different atomic weight.
     We have reviewed in \S 3.2 two distinct sites for
s-process nucleosynthesis when the $^{13}$C$(\alpha, n)^{16}$C
reaction is the neutron source: (1) a convective helium- and
carbon-rich shell (convective \nuc{13}{C} burning) during a
helium-shell flash and (2) a helium- and carbon-rich radiative
zone (radiative \nuc{13}{C} burning) during the interflash phase.  
     The convective site arises naturally in EMP model stars
within the framework of the current standard theory of stellar
evolution.  
     On the other hand, s-process nucleosynthesis in a radiative
site requires the formation of an appropriate \nuc{13}{C} pocket
by either semiconvective mixing (which has been shown to occur
in low-mass models of Pop.~II metallicity) or by convective
overshoot and/or by rotation induced mixing across a
carbon-hydrogen interface during the third dredge-up phase.  
From an analysis of the distribution of s-process elements
in the younger populations \citep[see the review by][]{bus99},
it has been concluded that the overshoot must occur.   
     The radiative \nuc{13}{C}-burning model (with the
\nuc{13}{C} made possible by convective overshoot) predicts that
the heavier s-process elements are produced at higher abundances
in stars of lower metallicity since the number of neutrons
available per seed nucleus increases with decreasing metallicity
unless the amount of $^{13}$C of the \nuc{13}{C} pocket
decreases proportionately.  

In Figure~7, the ratio of Pb to Ba, relative to solar, is
plotted against the stellar metallicity. Data are taken from
\citet{aok01,aok02b,joh02,coh03, luc03}.  
     The figure shows clearly that Pop.~II stars with
$\feoh > -2.5$ exhibit a large ${\rm Pb} / {\rm Ba}$ ratio,
of the order of ten times larger than the solar value, at least
qualitatively in agreement with the prediction of the radiative
\nuc{13}{C}-burning model.
     In addition, \citet{van01,van03} find, for CH stars with
$\feoh > -2.45$, enhancements by a factor of $\sim 10$ in the
ratio of Pb to the main s-process elements.

However, a sharp difference in the ${\rm Pb} / {\rm Ba}$ ratio
is discernible at the metallicity $\feoh \simeq -2.5$. 
     Just below this metallicity, with one exception, 
${\rm Pb} / {\rm Ba}$ ratios are smaller, by up
to a factor of 10 or so, than those above this
metallicity. 
     The exception is HE0024-2523, with $\feoh \simeq -2.7$
and $[\nucm{}{Pb} / \nucm{}{Ba}] \simeq 1.8$ \citep{luc03}.  
     This star is known to be in a binary system of
short period ($\sim 3$ days) and small semi-major axis
($\sim 10 R_\odot$).
     It is likely, as suggested by \citet{luc03}, that the
precursor of the present binary experienced a common envelope
event when the original primary became a TPAGB star and that
the common envelope evolution affected the nucleosynthesis in the
primary, leading to a distribution of s-process elements different
from that occurring in an isolated TPAGB star.    

The sudden decrease in the ${\rm Pb} / {\rm Ba}$ ratios at
$\feoh \simeq -2.5$ for stars which are not in a close binary
suggests that radiative \nuc{13}{C} burning in a \nuc{13}{C}-rich
pocket during a third dredge-up episode is either dominated by
or replaced by convective \nuc{13}{C} burning.
     In the convective \nuc{13}{C}-burning model, \nuc{13}{C}
is mixed into a helium convective zone which is much more massive
than the expected mass in the \nuc{13}{C} pocket
\citep[e.g., see][]{bus99}.  
     The mass of the mixed-in \nuc{13}{C} cannot be much larger
than assumed in the \nuc{13}{C} pocket \citep{gal88,gor00}
since the splitting of the convective zone places a ceiling
on the amount of \nuc{13}{C} produced and carried down into the
lower helium convective zone before the splitting occurs.  
     Accordingly, in the convective \nuc{13}{C}-burning model,
the number of available neutrons per seed nucleus can be smaller
compared with the number available in the radiative
\nuc{13}{C}-burning model, which may lead to a sudden decrease
in the $[{\rm Pb} / {\rm Ba}]$ ratio if there is a switchover
in the mechanism of s-process nucleosynthesis. 

Alternatively, \citet{aok02b} argue that, for metallicities
$\feoh \simeq -2.5$, the observations suggest a trend in the
[Pb/Ba] versus [Fe/H] diagram, with [Pb/Ba] decreasing with
decreasing $\feoh$.
     However, the suggested trend is spoiled by the
existence of the star CS 22183-015, which, at a metallicity
$\feoh \simeq -3.1$, has [Pb/Ba] much larger (\citet{joh02})
than that of most stars of metallicity $\feoh \sim -2.5 \pm 0.3$.
    The neutrons per seed made available in the
s-process nucleosynthesis event responsible for the abundances
in CS 22183-015 were larger by a factor of $3 \sim 4$
than those made available in the events responsible for
these abundances in stars with metallicities in the range
$-2.8 < \feoh < - 2.5$.
Also in the convective \nuc{13}{C}-burning model, the number of
available neutrons per seed nucleus increases in inverse
proportion to the abundance of iron-group elements, and this is,
at least qualitatively, in accord with the difference in the
$[{\rm Pb} / {\rm Ba}]$ ratio between CS 22183-015
and the stars with metallicities in the range $-2.8 < \feoh < - 2.5$.
In the binary scenario with convective \nuc{13}{C} burning,
dispersions in the ratios of s-process elements among individual
secondaries of similar metallicities may be ascribed
to variations in the amount of \nuc{13}{C} mixed into the
convective zone; this amount might be expected to vary with
the initial mass of the primary. In this interpretation, a
perceived trend in the element ratios for stars with metallicities
in the range $-2.8 < \feoh < - 2.5$ is instead dispersion
about a mean.

Finally, we note that there is no observational evidence
pointing to the operation of the radiative \nuc{13}{C}-burning
mechanism below $\feoh \simeq -2.5$.  
     Currently available observations are limited, and, certainly,
more observational data on the ${\rm Pb}/{\rm Ba}$ ratio,
especially for the lowest metallicities, are needed to help
clarify the situation and to allow definitive conclusions to
be drawn.  

Although the efficiency of the convective-overshoot
mechanism which prepares conditions for the formation of
a \nuc{13}{C} pocket has yet to be established as a function
of environmental conditions, it is natural to suppose that,
since the metallicity surely affects the thermal structure
near the boundary between the surface convective zone and
the carbon-rich helium core during the third dredge-up episode,
this efficiency may change with the metallicity.
     It is worth noting that, even for stars with $\feoh > -2.5$,
the observed ${\rm Pb} / {\rm Ba}$ ratios are 
small compared with those expected from s-process
nucleosynthesis models which rely on a neutron source which
does not depend on the metallicity.  
     In Figure 7 we plot the ${\rm Pb} / {\rm Ba}$ ratios predicted
by \citet{bus99} for a particular choice of
\nuc{13}{C}-pocket characteristics \citep{gali98}.  
     The predicted values all lie above the observed values
by more than a factor of $\sim 10$.
     This may again indicate that 
the strength of the neutron source in Pop.~I stars is larger
than the strength of the neutron source in Pop.~II stars.  
     One possibility is that the primary mechanism which sets
the stage for the formation of the \nuc{13}{C} pocket switches
from overshoot mixing to semiconvective mixing as the metallicity
drops below a critical value.  
     The semiconvective mixing mechanism has been demonstrated
to work for the metallicity $Z = 0.001$ \citep{hol89} and, due
to the character of the mixing, the ratio of protons to
\nuc{12}{C} in the mixed region can be smaller than expected
in the case of overshoot mixing, leading
ultimately to a smaller abundance of \nuc{13}{C} in the
\nuc{13}{C}-pocket.
     Further investigation is necessary to determine how
the conditions for the formation of a \nuc{13}{C}-pocket
vary with the metallicity.  For EMP stars which have experienced
the He-FDDM event, the opacities in radiative regions are
dominated by the carbon dredged-up into the envelope at the end of
this event, so semiconvection is not expected to play a role
in producing a \nuc{13}{C} pocket in them.

\section{The Origin of HE0107-5240}

In this section, we examine again the scenarios presented
in \S 2 for the origin of \mmps, this time taking advantage
of the theoretical and observational knowledge reviewed in \S 3.
     We focus first on scenarios that assume that the
observed carbon enrichment was acquired after birth.  
     \mmps\ exhibits large excesses of carbon and nitrogen and
${\rm C}/{\rm N} = 40 \sim 150$ \citep{chr02,chr04}. 
     Since single low mass Pop.~III stellar models develop
an abundance ratio
${\rm C}/{\rm N} \simeq 1$ as red giants \citep[Case I;
$ \le 1.1 \msun$][]{hol90,fuj00,sch02}, the possibility of an
evolved, single Pop.~III star is excluded.  
     Since \mmps\ is a red giant of metallicity less than
$\feoh \simeq -4.5$, the Case II model cannot be applied.

However, the Pop.~III binary scenario, with a primary
component of the Case II$^\prime$ variety, is a viable possibility.
Since model stars of metallicity $\feoh \lesssim -4.5$ share
common evolutionary characteristics, the second generation binary
scenario, with metallicity at birth being the same as presently
observed, is also a possibility. Accordingly, we postulate in the
following that HE0107-5240 is either a first or second
generation star, formed out of matter in a primordial cloud, and
formed in a binary system in which it was the secondary of
mass $M_{\rm s} \simeq 0.8 \msun$ and in which the primary
was of initial mass in the range $1.2 \leq M_{\rm p} / \msun \leq 3.0$. 
     The primary component has followed Case II$^\prime$ evolution
and achieved enrichment of C, N, O, and Na by experiencing
a He-FDDM event and subsequent third dredge-up events, as discussed
in \S 3.  

Figure~4 shows the element
abundances resulting from neutron-capture nucleosynthesis
in the helium-flash convective zone during a He-FDDM event
for a model in which the abundance of \nuc{13}{C} mixed into
the helium-flash driven convective zone,
relative to the abundance of \nuc{12}{C} in the
zone, is $\nucm{13}{C}/\nucm{12}{C} = 0.001$.   
     The abundances of \nuc{16}{O} and of Mg and Na isotopes
have reached their final values at the right hand side of the
figure.
     Along the right hand ordinate in the figure are placed
the abundances relative to carbon of O (open circles),
Mg (crosses), and Na (filled circles) observed for \mmps. 
     For O, Na and also Mg, the observed ratios can be well
reproduced by the calculated abundance ratios to carbon in
the helium convective zone.  
      As discussed in \S 3.3, the abundances of C and O as well as Na, achieved during a
He-FDDM episode are not expected to be changed much in
the helium-flash convective
zones which appear during the following TPAGB phase.  
     In Fig.~5, the shaded region gives the range in the
${\rm C} / {\rm O}$ abundance ratio estimated from the
observations of \mmps. 
     An abundance ratio in the observed range is achieved
during helium shell flashes in model stars of mass $M \le 3 \msun$
if temperatures near $T \simeq 2.3 \times 10^8$ K are not
reached.  

The surface enrichment of carbon in the primary component
depends upon what fraction of the matter once contained 
in the helium-flash driven convective zone during the He-FDDM
episode and the following third dredge-up episode is carried
outward into the convective envelope.  
     Denoting the mass of dredged-up material by
$\Delta M_{\rm d u}$, we have
the surface carbon abundance, $X_{12, \rm p, env}$, in the envelope of the primary by mass at 
\begin{equation} 
     X_{12, \rm p, env} = X_{12, \odot} \left({\Delta M_{\rm d u} \over 0.03\ \msun }\right)\ \left({1.5 \msun \over M_{\rm p, env}}\right)\ \left({X_{12, \rm p, He} \over 0.15 }\right), 
\end{equation} 
     where $M_{\rm p, env}$ is the mass in the envelope of the primary,
$X_{12, \rm p, He}$ is the carbon
abundance by mass in the helium convective zone of the primary, and
$X_{12, \odot}$ is the solar carbon abundance by mass
\citep[$= 3.0 \times 10^{-3}$;][]{and89}.
The mass in the helium-flash driven convective zone
falls in the range
$\simeq {\rm a \ few} \times 10^{-2}
 \sim {\rm several} \times 10^{-3} \msun$,
and eq.~(8) indicates that only a small fraction of this,
or $\Delta M_{\rm d u} 
\sim 3.5 \times 10^{-5} M_{\rm p, env} X_{\rm 12, p, He}^{-1}$ is
enough to enrich the envelope above $[{\rm C N O}/{\rm H}] = -2.5$. 
As a corollary, we note that this is sufficient to prevent
another He-FDDM event from occurring. 

Immediately after the He-FDDM episode, the C/N ratio
in the envelope of the primary is $\sim 5$. Subsequently,
the C/N ratio increases as \nuc{12}{C} is added to the envelope
during successive third dredge-up events. 
     In order for the ${\rm C}/{\rm N}$ ratio to reach the
observed range, 8-30 times as much carbon has to be added by
third dredge-up episodes as is added by the He-FDDM episode.
     If approximately $\hbox{a few} \times 10^{-3} \msun$ is added
by the He-FDDM episode, third dredge-up episodes will cause
the C/H ratio at the surface of the primary to approach, or
even exceed, the solar C/H ratio.
     The $\nucm{12}{C}/\nucm{13}{C}$ ratio at the surface of
the primary at the beginning of the mass-transfer event can
be estimated if we assume that, during the He-FDDM episode, the
$\nucm{12}{C}/\nucm{13}{C}$ ratio in the hydrogen-flash
convective region attains the equilibrium value of $\sim 4$.
Dredge-up following the He-FDDM event leads to a similar carbon
isotopic abundance ratio at the surface. Then, 8-30 subsequent
third dredge-up episodes (during which only the isotope
$\nucm{12}{C}$ is added) lead to a surface ratio
$\nucm{12}{C}/\nucm{13}{C} \sim 32-120$, which is consistent
with the estimate for \mmps\ of
$\nucm{12}{C}/\nucm{13}{C} > 50$ \citep{chr04}.

As the enhancement of carbon in its envelope progresses,
the primary brightens and expands in response to the
increase in the abundance of CNO catalysts and 
the corresponding increase in the burning rate of its
hydrogen-burning shell.
     Finally, the primary loses its hydrogen-rich envelope
either through Roche-lobe overflow or through wind mass loss;
in either case, some envelope matter falls onto the secondary.  
     The amount of mass that must be transferred from the
primary to the secondary to produce the surface carbon
abundance $[{\rm C}/{\rm H}] = -1.3$ observed for \mmps\ may
be estimated as
\begin{equation}
     \Delta M_{\rm acc} = 0.01\ \msun \ 
      \left({X_{12, \odot}\over X_{12, \rm p, env}}\right)\
          \left({M_{\rm s, c\ env}\over 0.2 \msun}\right),   
\end{equation}  
     where $M_{\rm s, c\ env}$ is the mass
of the envelope convective zone of the secondary after
mass transfer ceases.  
     The fact that $\Delta M_{\rm acc}$ is only a small fraction
of the initial envelope mass of the primary suggests that mass
transfer occurs via accretion from a wind rather than through
Roche-lobe overflow, in agreement with the analysis given by
\citet{ibe00}.  
     In the case of accretion from a wind, the mass-accretion
rate onto the secondary may be estimated from the gravitational
focusing cross section defined by eqs.~(\ref{gfcrosssection})
and~(\ref{impact}).   
     For a spherically symmetric wind, the mass
$\Delta M_{\rm acc}$ accreted by the secondary is simply the
mass $\Delta M_{\rm loss}$ lost by the primary times the solid
angle formed by the gravitational focusing cross section
at a distance equal to the separation $a$ of the binary
components. Thus,
\begin{equation}
                         \Delta M_{\rm acc} =
           {\sigma_{\rm acc} \over 4 \pi a^2}\ \Delta M_{\rm loss}\ .
\end{equation} 
     From this relationship, we may estimate the semi-major
axis of the binary system at the start of mass transfer as
\begin{equation}
a \simeq 18 \hbox{ AU}\ \left({M_{\rm s}\over  0.8 \msun}\right)\
\left({0.01 \Delta M_{\rm loss}\over \Delta M_{\rm acc}}\right)^{1/2}
   \left({20\ \hbox{km s}^{-1}\over v_{\rm rel}}\right)^{2},
\end{equation} 
where $v_{\rm rel}$ is velocity of the wind relative to the
velocity of the secondary (the vector difference of orbital
velocity and wind velocity).
     If the wind carries away matter with the same specific angular
momentum as resides in the orbital motion of the primary, the
orbital separation increases by a factor inversely
proportional to the total mass of the system (Jean's theorem).
     Accordingly, the resultant orbital period $P_{\rm orb}$ of 
the system at the end of mass transfer could be of the
order of
\begin{equation}  
                P_{\rm orb} \simeq 76 \hbox{ years} \
         \left({M_{\rm total, 0}^{3/2} \over M_{\rm total}^{2}}\right) \
            \left({M_{\rm s} \over 0.8 \msun}\right)^{3/2}
                    \left({0.01 \Delta M_{\rm loss}
            \over \Delta M_{\rm acc} } \right)^{3/4}\
        \left({20\ \hbox{km s}^{-1}\over v_{\rm rel}}\right)^{3},
\end{equation} 
     where $M_{\rm total, 0}$ and $M_{\rm total}$ are, respectively,
the total mass of the binary before and after the completion
of mass transfer.

Since the wind escapes the primary with a speed of the
order of the sound speed at the surface of the primary
($\sim 10 \hbox{ km s}^{-1}$) and the orbital speed of the
secondary is of comparable magnitude, we may estimate
$v_{\rm rel} \sim 20$ km s$^{-1}$.  
     Thus, after mass transfer is completed, the separation
and period of the binary could be increased up to 
$ a \simeq 34$ AU and $P _{\rm orb} \simeq 150$ years
with $M_{\rm total} \simeq 1.5 \msun$
and $M_{\rm total, 0} \simeq 2.8 \msun$.
     Such a long period is not excluded by the observations,
which, to date, cover only 52 days in one case \citep{chr04} and
373 days in another \citep{bes04}.
 The above estimate predicts a variation in the velocity, $v_r$, in the radial direction at an order of $d v_r / d t \simeq 0.11 \sin i \sin \theta \hbox{ km sec}^{-1} \hbox{ yr}^{-1}$, where $i$ and $\theta$ are the inclination angle and the orbital position angle, respectively.  
     This remains well below the $1 \sigma$ scatter of the radial velocity measurement of \mmps, which is currently $44.5 \pm 0.5$ km s$^{-1}$ \citep{chr04}.

The argument concerning abundances which we have presented
applies to a binary system of either the first or second
generation, independent of whether the iron-group elements
are acquired after birth or are present at birth.  
     With regard to s-process elements, however, it does
make a difference whether or not seed nuclei exist
in the helium-flash convective zone.  
     If \mmps\ is a second-generation star, the s-process
elements are synthesized with iron-group elements
as seed nuclei in the lower helium convective shell during
the He-FDDM episode.  
     Since the iron abundance of HE0107-5240 is smaller by
a factor of more than $\sim 100$ than the iron abundances
of most other known EMP stars, the number of neutrons available
per seed nucleus ought to be larger in \mmps\ by the same factor.  
     Even if convective \nuc{13}{C} burning is responsible for
the synthesis of heavy s-process elements, 
very large ${\rm Pb}/{\rm Ba}$ ratios would be expected.  
     As most of the iron-group elements are converted into
heavy s-process elements, the overabundance of lead in the
helium-flash convective zone is given by the number ratio
of Fe to Pb in the solar distribution, namely
 $(Y_{\rm Fe}/Y_{\rm Pb})_\odot = 3 \times 10^5$.
     In the binary scenario, the s-process elements produced
in the He-FDDM event are first diluted by mixing into the
convective envelope of the AGB primary and then, after transfer
via a wind to the secondary, by mixing into the convective
envelope of the secondary when it becomes a red giant.
Hence, the overabundance at the surface of \mmps\ becomes
\begin{eqnarray}
         [{\rm Pb}/{\rm Fe}] & \simeq & 
  \log \left[ \left({Y_{\rm Fe}\over Y_{\rm Pb}}\right)_\odot\ 
     \left({f \Delta M_{\rm d u} \over M_{\rm p, env}}\right)\
  \left({\Delta M_{\rm acc} \over M_{\rm s, c, env}}\right) \right] \\   
 & \simeq & 2.5 + \ \log \left[\ f \
 \left({0.15 \over X_{\rm 12,p,He}}\right)\right],
\end{eqnarray}   
     where $\Delta M_{\rm d u}$ is the mass of dredged-up material
and $f$ is the fraction of this dredged-up material
which has been exposed to neutron irradiation during the He-FDDM
event. 
    Since a significant fraction $r$ of matter is incorporated
into helium-flash convective zones during subsequent thermal pulses,
$f = \sum^{8 \sim 30}_{n}  r^n /(8 \sim 30)
 \simeq 0.2 \sim 0.05$ for $r \simeq 0.6$ \citep{ibe77}.  
     Consequently, an overabundance of lead
($[{\rm Pb}/{\rm Fe}] \simeq 1 \sim 2$) ought to be observed
for \mmps\ if it really is a second generation star.

On the other hand, if \mmps\ is a first generation star, two
different pictures are possible, depending on whether or not
accreted metals can be carried into the helium convective zone
during the He-FDDM episode.
     The primary component may accrete metal-polluted gas in the
primordial gas cloud, and carry these metals inward as surface
convection deepens at the time of the second dredge-up phase.
     In Figure~8, the dotted line shows the deepest reach of
surface convection during the second dredge-up phase as a function
of the mass of a Pop.~III model star in the mass range
$1.2 \le M/\msun \le 3.0$. In real counterparts of the model stars,
metals accreted onto surface layers are spread uniformly
throughout the shaded area shown in the figure.
Solid lines show the location of the base and outer edge
of the convective shell formed in the He-FDDM event at the
moment that the outer edge touches the base of the hydrogen
profile.

There is a critical mass, $M_{\rm crit}  \simeq 2.5 \msun$,
such that, for an initial mass $M_{\rm initial} < M_{\rm crit}$,
the He-FDDM convective shell remains below the polluted (shaded)
area and heavy s-process elements built-up from iron-group seed
nuclei will not be produced.  
     In contrast, for an initial mass $M_{\rm initial} > M_{\rm crit}$,
the He-FDDM convective shell extends into or lies completely in the
polluted (shaded) area and heavy s-process elements are built-up
from iron-group seed nuclei.
     Penetration into the polluted area has been made possible
by the fact that, following a second dredge-up episode and
prior to the He-FDDM episode, hydrogen burning during the
EAGB phase of evolution has moved the hydrogen-helium interface
into the metal-polluted zone.
     In this case, however, the degree of enhancement varies with
the degree of metal contamination that the primary has experienced
during its relatively short lifetime in the polluted primordial
cloud. On average, a less massive star
(lifetime $t\sim 4.1 \times 10^8$ yr for $M_{\rm initial} = 2.5 \msun$)
will experience more contamination than a more massive star
(lifetime $t\sim 2.6 \times 10^8$ yr for $M_{\rm initial} = 3.0 \msun$)

Our results for the binary scenario are summarized in Figure~9.  
     Our proposal is that \mmps\ has evolved from a 
secondary component (of mass $\sim 0.8 \msun$) in a wide binary
system with a primary (of initial mass between $1.2 \msun$ and
$3 \msun$) which has undergone Case II$^{\prime}$ evolution and
thereafter transferred via a wind a small amount of mass to the
secondary.  
     The lower boundary of the mass range in Fig.~9 is determined
by the condition that the third dredge-up takes place after
the surface CNO abundance grows larger than
$[{\rm C N O}/{\rm H}] \gtrsim -2.5$.  
     Our binary scenario gives a reasonable account of the
observed abundances of C, N, O, and Na, and of the carbon isotopic
ratio $\nucm{12}{C}/\nucm{13}{C}$, independently of whether
the binary system is composed of first-generation stars which
accrete metals after birth or of second-generation stars,
formed of gas which has already been contaminated with metals.

Clues to the origin of an observed metallicity can be obtained
by observing the relative abundances of heavy s-process elements.  
     If, for example, $[{\rm Pb} / {\rm Ba}] \sim 0$, \mmps\ may
be identified as a Pop.~III star born in a binary system with
a primary component of mass $M_{\rm initial} < M_{\rm crit}
(\simeq 2.5 \msun)$.  
     If, on the other hand, $[{\rm Pb} / {\rm Fe}] = 1 \sim 2$,
\mmps\ may be identified as possibly being a second generation star.  
     A Pb overabundance is also consistent with the primary being a
Pop.~III star with $M_{\rm crit} < M_{\rm initial} \lesssim 3 \msun$,
but the degree of enhancement depends upon the efficiency
of metal accretion in the primordial cloud.  
     Presently, we have only upper limits on the abundances
of heavy s-process elements: $[{\rm Sr}/{\rm Fe}] < -0.52$,
 $[{\rm Ba}/{\rm Fe}] < 0.82$,  $[{\rm Eu}/{\rm Fe}] < 2.78$
\citep{chr02,chr04}.  
     These limits are consistent with our models, but not
definitive confirmation.
In order to establish the origin of \mmps\ unambiguously,
we need abundances rather than limits for these ratios and, most
importantly, we require the ratio ${\rm Pb} / {\rm Ba}$ 

Finally, we follow the evolution to the red giant branch
of a $0.8 \msun$ model star with the C and O abundances
observed for \mmps\ and investigate the modifications of
the surface abundances along the RGB.  
     We add \nuc{22}{Ne} to see the effect on the
\nuc{23}{Na} abundance.  
     Figure~10 shows the abundance distribution in the envelope
initially (dotted lines) and sometime after the first dredge-up
event has taken place (solid lines).  
     At its maximum inward extent, surface convection reaches
$M_r = 0.3387 \msun$ when the core mass is $M_1 = 0.2645 \msun$
(mixing length = 1.5 times the pressure scale height).
     This deepest reach is much deeper than in a pure Pop.~III model
where, at its maximum inward extent, surface convection reaches
only to $M_r = 0.5903 \msun$ when the core mass is $M_1 = 0.3454 \msun$. 
As a result of the first dredge-up event, the surface nitrogen
abundance increases to $X_{\rm N} = 1.6 \times 10^{-6}$ at the
expense of carbon, leading to the ratio $\nucm{}{C} / \nucm{}{N}
\simeq 100$, which is in the range consistent with the observations
for \mmps.  
     But \nuc{22}{Ne} hardly burns in layers which can be reached
by the base of the convective envelope during the first dredge-up
episode, and, hence, Na cannot be produced during
evolution along the RGB phase of an isolated single star.
     One also cannot invoke extra mixing of the sort implied
by abundances observed for some giants in globular clusters
which are known to exhibit surface enhancements of sodium; such
mixing brings about the depletion of carbon and the enhancement
of nitrogen at the same time that it brings about enhancement
of sodium.
     This may constitute an additional argument against the single
star scenario of \citet{ume03} (see \S 2), which assumes birth
out of a small cloud polluted with the ejectum of a supernova with
an unusual carbon enrichment.

\section{Conclusions and Discussion} 

During the past decade, there has been considerable observational
and theoretical progress in understanding the properties of
extremely metal-poor (EMP) stars in the Galactic halo.
     Thanks to the HK survey \citep{bee92}, the number of known
EMP stars ($\feoh \lesssim -2.5$) has reached more than 100,
and the spectroscopic characteristics of these stars have
been revealed through detailed studies using large telescopes.  
     By analyzing existing theoretical and observational
evidence and by utilizing the results of new computations,
we have focused in this paper on the peculiarities that distinguish
EMP stars from Pop.~I and II stars and have presented a general
theoretical framework for understanding the evolution of EMP
stars and for understanding the nucleosynthesis which has taken
place in their interiors. Our framework relies heavily on a binary
scenario in which the primary has produced, and transferred to
the secondary, many of the isotopes which are present at the
surface of an observed EMP star.

A distinct feature that characterizes the evolution of EMP stars
of low and intermediate mass is the helium-flash driven deep mixing
(He-FDDM) phenomenon, which is triggered when the outer edge of
a convective zone driven by a first helium-shell flash (which occurs
near the beginning of the AGB phase in intermediate mass models and
at the tip of the RGB in low mass models) extends into
hydrogen-rich material.
     The first discussions of the He-FDDM mechanism
\citep{fuj90,hol90,fuj95,fuj00} drew attention
to the surface enhancement of carbon and nitrogen made
possible by this mechanism.  
     In the present paper, we have investigated another consequence
of the He-FDDM event, namely, the nucleosynthesis of s-process
elements which occurs when neutrons are released by the reaction
$\nucm{13}{C} (\alpha, n) \nucm{16}{O}$ in the helium- and
carbon-rich convective zone.
     We demonstrate that products of this nucleosynthesis can
lead to surface enrichments of O, Ne, Na, and Mg and, if iron-group
seed nuclei are present in the convective zone, to surface enrichments
of heavy s-process elements.  
     In addition, stars which evolve to the TPAGB phase
after having experienced a He-FDDM event can also develop a
large $\nucm{12}{C}/\nucm{13}{C}$ ratio at the surface in
consequence of third dredge-up events.

Our binary scenario gives a reasonable account
of the observed properties of EMP stars, such as the very high
frequency of carbon-rich stars and s-process nucleosynthesis
products which differ from those made by stars of younger
populations.  
     The scenario enables us in principle to identify
modifications in surface abundances which, after birth in an
unpolluted primordial cloud, first generation EMP stars may
have experienced due to accretion of primordial matter polluted
by the ejecta of first generation supernovae; this ability is
essential if we wish to use low mass EMP star survivors as
tools to probe the early Universe.

We have constructed a specific binary scenario to account
for the observed abundance characteristics of \mmps.
The initial binary system has the properties:
separation $\sim 18$ AU, orbital period $\sim 45$ years,
a secondary of mass $\sim 0.8 \msun$, and a primary of mass
in the range $1.2 \lesssim M / \msun \lesssim 3$.  
     The primary follows an evolutionary path of the Case
II$^\prime$ variety in the classification scheme defined by
\citet{fuj00}. In consequence
of experiencing a He-FDDM episode, the primary develops surface
enhancements of C and N and,
in consequence of experiencing third dredge-up events during
the subsequent TPAGB phase, develops a large overabundance of C
relative to N as well as enhancements of O, Ne, and Na, formed during the He-FDDM and subsequent thermal pulses.  
     After ejecting its hydrogen-rich envelope in a superwind,
the primary evolves into a white dwarf. The secondary accretes
heavy-element enriched matter from the wind emitted by the primary
and, when it evolves into a red giant, it mixes this enriched
matter into a deep convective envelope, establishing surface
abundance peculiarities similar to those observed for \mmps.

Because the iron abundance is so small ($\feoh=-5.3$), the
presence of iron-group elements in \mmps\ can be attributed to
either (1) accretion after birth of gas in a parent primordial cloud
which has been polluted with material ejected by one or more
first generation supernovae or (2) birth out of already
polluted matter in the parent cloud.
     
We have shown that further light on the source of iron-group
elements can be shed by comparing an observed distribution of
heavy s-process elements with theoretical expectations. 
     An abundance ratio $[{\rm Pb}/ {\rm Fe}] \simeq 1 \sim 2$,
coupled with a large ${\rm Pb} / {\rm Ba}$ ratio would be evidence
that \mmps\ is a second generation star, or a Pop.~III star in a
binary with a primary of initial mass in the range
$2.5 \lesssim M / \msun \lesssim 3.0$.  
     A lack of heavy s-process element enrichment would
indicate that \mmps\ is really a Pop.~III star in a binary
with a primary component of initial mass in the range
$1.2 \msun \lesssim M \lesssim 2.5 \msun$.  
     Unfortunately, current observations provide only upper
limits on the abundances of light and main-line s-process
elements, so no conclusion can as yet be drawn.
     A ratio $[{\rm Pb} / {\rm Fe}] \simeq 1 \sim 2$ translates
into a ratio $[{\rm Pb} / {\rm H}] \simeq -4 \sim -3$ for \mmps.
This is difficult to detect with present facilities, and we will
probably have to wait for future observations with a next generation
large telescope and the elaboration of model atmosphere including,
for example, 3D hydrodynamic simulations to establish definitively
whether or not HE0107-5240 is a Pop.~III star.

With regard to the current binary status of \mmps, variations in
the radial velocity of the size predicted by our scenario cannot be
excluded by extant spectroscopic observations which, to date,
cover only 52 days \citep{chr04} and 373 days \citep{bes04}.
     Because of wind mass loss, the initial binary system may
have been considerably widened. Adopting Jean's theorem which
predicts the constancy of the product, semi-major axis times
the total mass, we would expect the current binary to have the
characteristics $a \simeq 34$ AU and $P_{\rm orb} \simeq 150$
years, giving an orbital velocity of $\sim 7 \hbox{ km s}^{-1}$.
Confirmation of such a small velocity demands long term
observations at high dispersion.

As far as alternative single star interpretations of \mmps\
are concerned, any viable scenario must begin with formation
out of gas with a singularly unusual abundance distribution
(including the currently observed carbon and oxygen enhancements)
of the sort not expected by mixing products of normal supernovae
with primordial matter. 
     As an example, a scenario proposed by
\citet{ume03} supposes that the gas out of which \mmps\ was formed
acquired a carbon abundance of $[{\rm C}/{\rm H}] = -1.3$ after
the mixing of an unusually small amount of primordial matter
with a supernova ejectum in which the mass of carbon is
$\simeq 0.2 \msun$. 
     Given that a typical type II supernova ejects
a mass of the order of $0.1\sim 1 \msun$ in the form of Fe,
a most peculiar process of star formation is required to account
at the same time for a ratio of $\feoh \simeq -3$ characteristic
of EMP stars.
     It is to be noted that the mass of carbon in the supernova
ejectum cannot be much larger than the mass of iron in the ejectum
since, in order to obtain the observed ${\rm C} /{\rm O}$ ratio,
the carbon can only have come from the shell of partial helium
burning existing above the region where iron-group elements are
formed.  
     In addition, since, in contrast with nitrogen, which can
be produced in the interior and brought to the surface during
the first dredge-up phase on the RGB, sodium cannot be formed
during the evolution of a low mass star, and a non-canonical
mechanism for sodium enrichment has to be invoked.
     Given this hurdle for the (any) single star scenario, it
seems reasonable to accept the observed Na enhancement as evidence
that some of the matter in the convective envelope of \mmps\
was formed in the interior of a primary companion during the
AGB phase.

Although there is a controversy as to whether or not a single
star of mass as small as that of \mmps\ can be formed in a
primordial cloud, a condensation as massive as the
$2 \sim 3.8 \msun$ predicted by our binary scenario is not
excluded by current theory \citep{nak01}, which suggests a
bi-modal star-formation mass function for first generation stars.
     The initial condensation could be formed as a first generation
object in the first collapsed, primordial cloud of total (dark and
baryonic) mass $\sim 10^6 \msun$.  
     Another possible site is a primordial cloud of mass
$\gtrsim 10^8 \msun$, for which virial temperatures are
higher than $10^4$ K.  
     In such a cloud, \mmps\ could have been born as a first
generation star, but the formation epoch would be delayed in
comparison with the formation epoch of first generation stars in
a lower mass collapsed cloud. 

If \mmps\ is a second generation star, it also must have been
formed in a cloud of total mass $ \gtrsim 10^8 \msun$.
     In a primordial cloud of total mass $ \sim 10^6 \msun$ and
baryonic mass $\sim 10^5 \msun$, contamination of primordial gas 
with the $0.1 \sim 1 \msun$ of iron ejected by a typical supernova
would produce an iron abundance $\feoh = -3 \sim -2$.
     This metallicity is appropriate for most EMP stars known to date,
but it is hundreds of times larger than the metallicity of HE0107-5240.
     In order to achieve a metallicity appropriate for \mmps,
the contaminating supernova must eject an abnormally small mass
of iron.
     But, the supernova explosion would be so weak that the remnant
would dissolve into the interstellar gas instead of compressing
it to collapse conditions, and star formation would not be
triggered \citep{mac04}. 

As described in the introduction, the history of the search for
extremely metal-poor stars has taught us that, as the limiting
magnitude of a survey is increased, stars of lower metallicity
are detected, suggesting that, for EMP stars, there may be a
correlation between typical metallicity and apparent luminosity,
and hence, a relationship between typical metallicity and spatial
distribution.
     In the current framework of structure formation, galaxies
 have been formed by the merging of lower-mass building blocks.    
     If the parent cloud of \mmps\ is of mass $ \gtrsim 10^8 \msun$,
it differs from parent clouds of mass $\simeq 10^6 \msun$ out
of which many second generation stars with
$\feoh \simeq -4 \sim -2.5$ may have been formed.
     Accordingly, in addition to commenting on initial abundances
in primordial matter and characteristics of nucleosynthesis
in supernova explosions in the early Universe, EMP stars may 
also reveal the masses and locations in our Galaxy of their parent
clouds. 

The discovery of \mmps\ has also demonstrated the importance of
probing for Pop.~III stars by sorting according to carbon-star
characteristics rather than according to the weakness and/or
absence of the Ca II K line.
     As pointed out by FII00, it makes sense to search for
carbon stars since low-mass stars of $\feoh \lesssim -4.5$ spend
their final nuclear burning lives as luminous carbon stars
on the horizontal and asymptotic giant branches.  
     Assuming that it has the luminosity of a typical red giant,
\mmps\ is at a distance of $\sim 10$ kpc.  
     Because the luminosity of a typical intermediate mass
TPAGB star is larger by about a factor of ten than that of a
typical low mass RGB star, TPAGB stars are detectable at much
larger limiting magnitudes than are RGB stars.
     The down side is that the lifetime of a typical TPAGB star
is over ten times smaller than that of a typical RGB star.
Nevertheless, with CCD cameras, larger telescopes, and patience,
we may be able to find Pop.~III carbon stars further out in
the Galactic halo and perhaps even in intergalactic space.  

Recently, \citet{mar02} have reported a number of faint high
latitude carbon stars from the data obtained by the Sloan Digital
Sky Survey (SDSS). 
     Among these stars there may be EMP carbon stars of the sort
that we suggest looking for, even though \citet{mar02} argue that
nearby dwarf carbon stars outnumber giants in their sample.  
     Follow-up spectroscopic observations may decide the issue.
     Since their selection is based on comparisons with known types
of carbon stars in the five color system of SDSS \citep{kri98},
there is the possibility that Pop.~III carbon stars may be missed.
Because of different atmospheric properties related to higher
surface temperatures and different chemistry, the color properties
of EMP carbon stars may differ significantly from those
of more familiar carbon stars of Pop.~I, or even from those of EMP
carbon stars with $\feoh \simeq -3$.  
     In any case, it is important to continue the search for
carbon stars of low metallicities using every stratagem that
can be devised, including methods that make use of spectral lines
for which carbon molecules are responsible.  
    From extremely metal-poor carbon stars found at extreme
distances, or even from the absence of such stars, we may learn
important information about the early Universe.   

\acknowledgements

We thank the referee, Norbert Christlieb, for valuable comments
about the abundance of lead and for bringing to our attention
the importance of predicting the carbon isotopic ratio.
      We also thank Drs. W. Aoki, N. Iwamoto, K. Nomoto, Y. Yoshii, 
and T. C. Beers for valuable discussions.  
      This paper is based on one of the author's (TS) dissertation
submitted to Hokkaido University, in partial fulfillment of
the requirement for the doctorate.
      This work is in part supported by a Grant-in-Aid for
Science Research from the Japanese Society for the Promotion
of Science (grant no. 15204010).   
      One of us (II) thanks the Japanese Society for the
Promotion of Science for an eminent scientist award, permitting
an extended visit to Hokkaido University.

\clearpage

\begin{figure}
\plotone{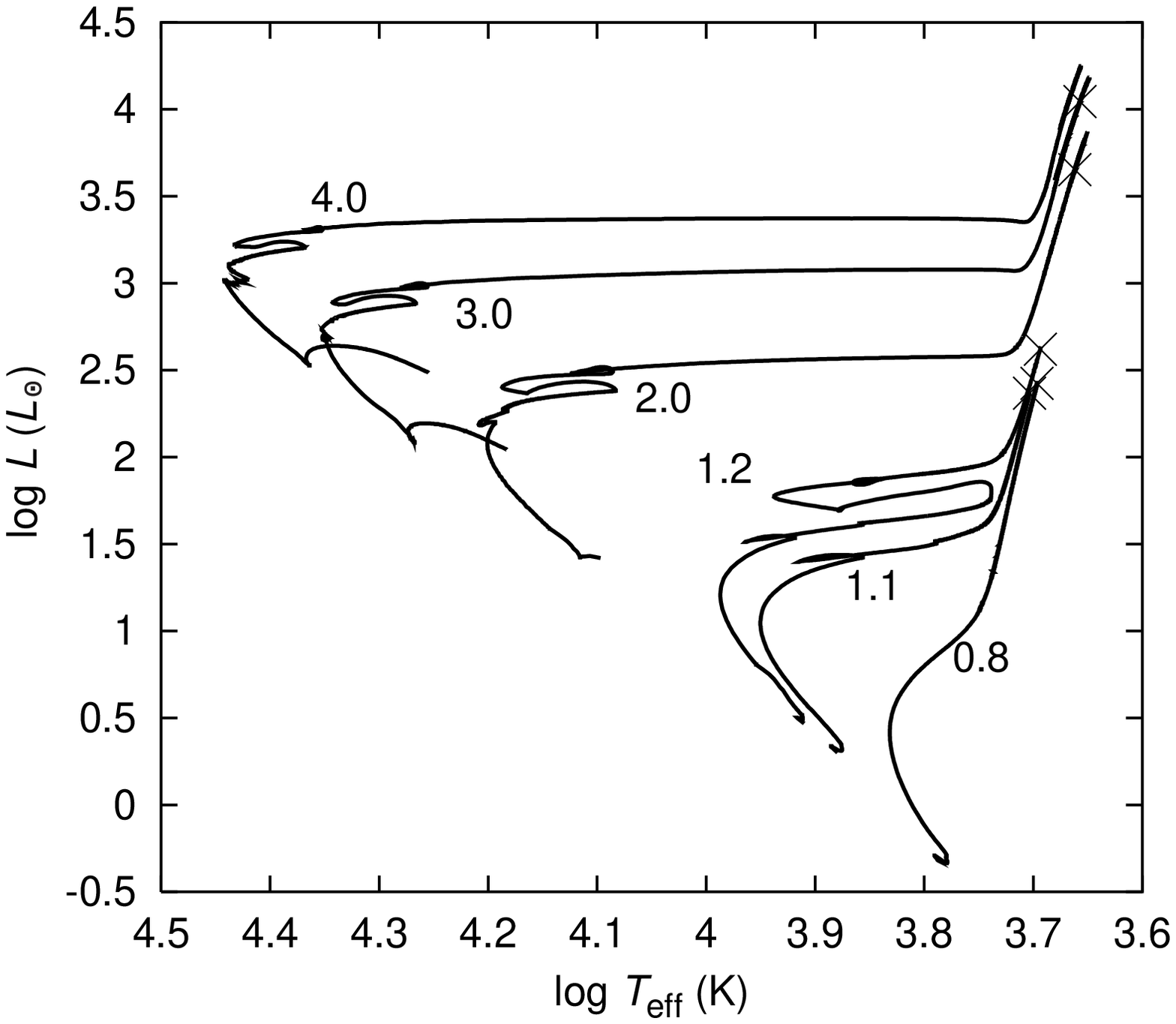}
\caption{
Evolutionary trajectories of Pop.~III model stars in the H-R diagram. 
     Numbers beside each curve denote the model mass, $M$, in units of
the solar mass.  
     For model stars of mass $M \le 3 \msun$, evolution is
followed from the zero age main sequence through the stage when
the helium-flash driven convective zone engulfs hydrogen and initiates
the helium-flash driven deep mixing (He-FDDM) mechanism.
     Evolution of the model of mass 4 $\msun$ is terminated 
along the TPAGB branch when thermal pulses reach an asymptotic
strength.
     Crosses mark where the He-FDDM event begins, on the
red giant branch for models of mass $M \leq 1.1 \msun$, or
at the beginning of the TPAGB for models of mass
$M \geq 1.2 \msun$. 
}
\end{figure}

\begin{figure}
\plotone{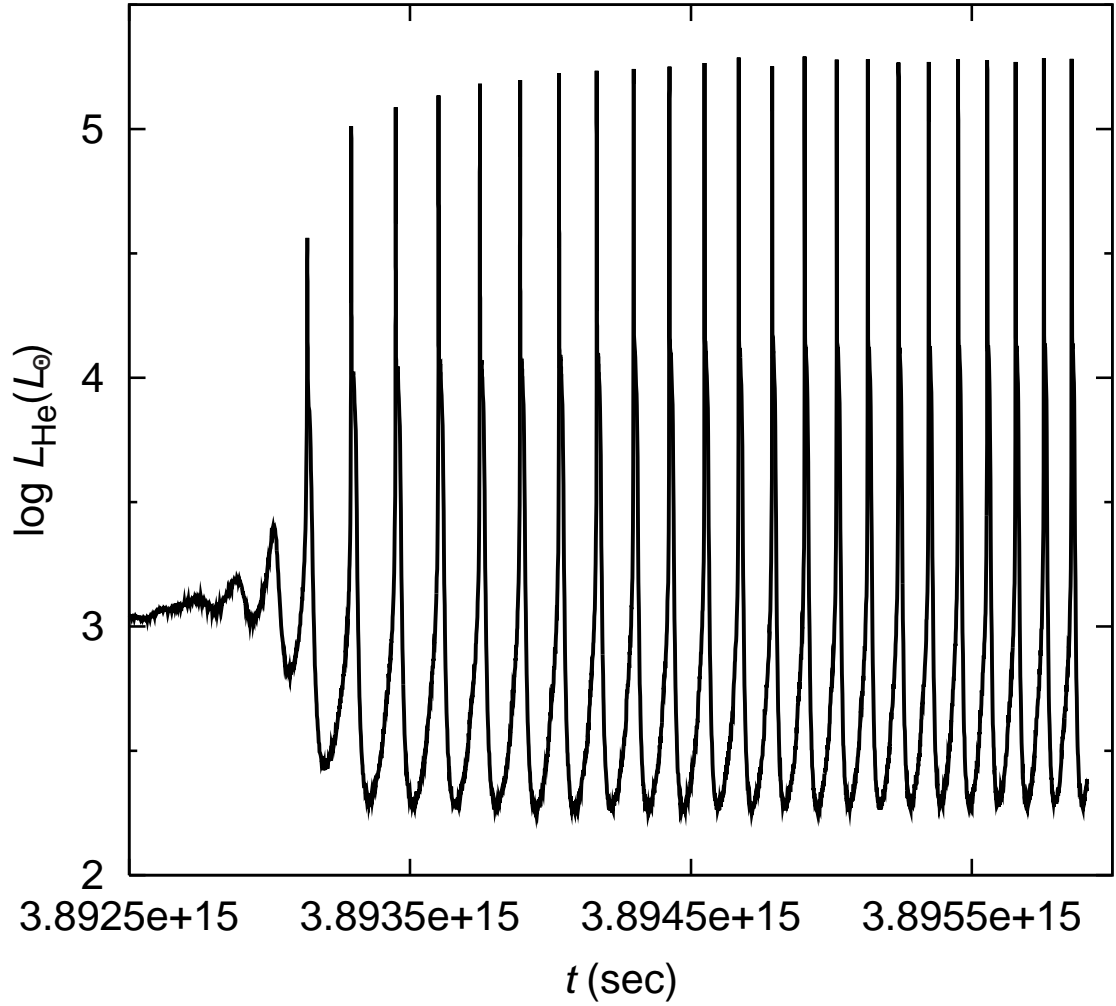}
\caption{
Variations in the helium-burning luminosity during the TPAGB phase
for a Pop.~III star of mass $4 \msun$. 
}
\end{figure}

\begin{figure}
\epsscale{.80}
\plotone{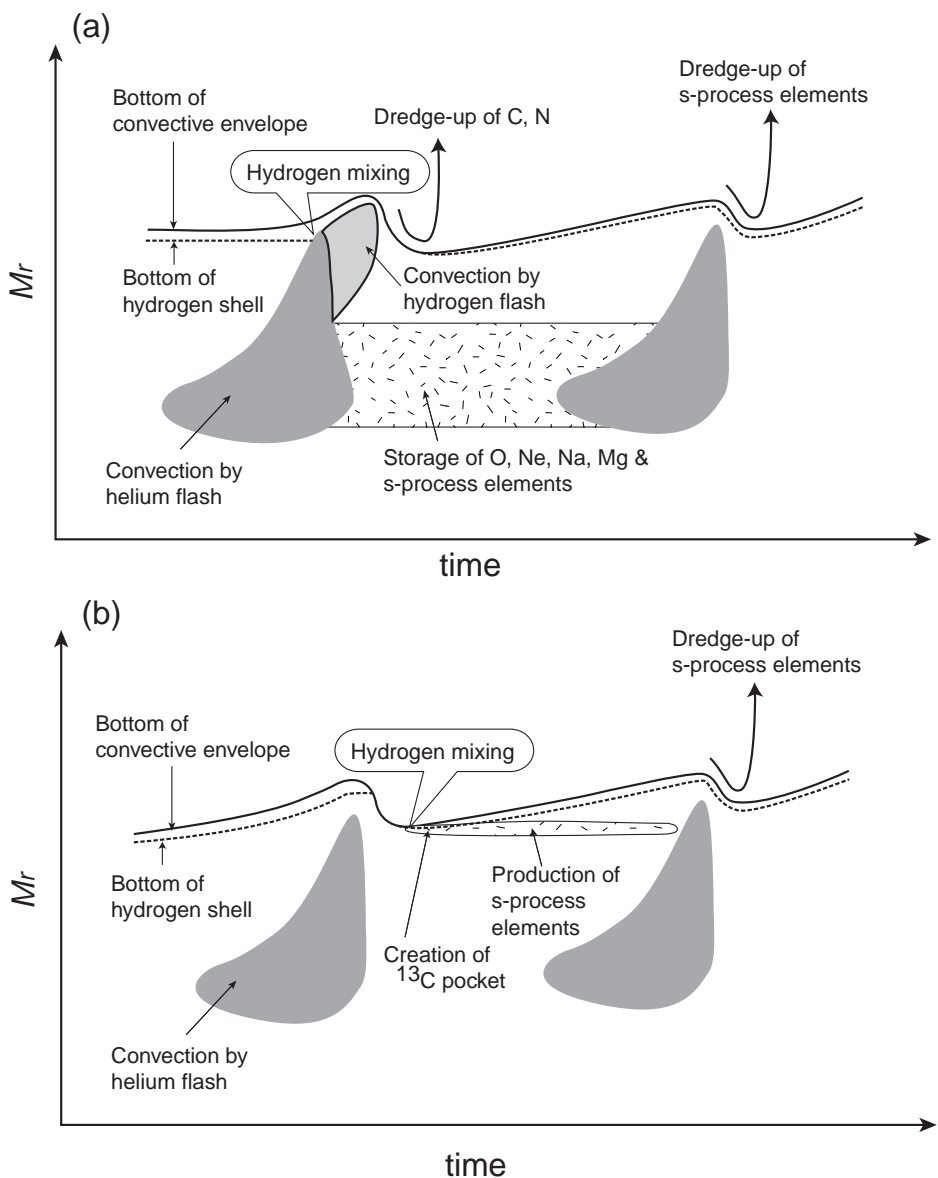}
\caption{
Schematic representations of the time dependence of (1) convective
zones associated with helium-shell flashes and (2) sites for the
nucleosynthesis and/or storage of s-process elements associated
with these flashes.
     Panel (a) describes characteristics of the convective
\nuc{13}{C}-burning model associated with a He-FDDM episode.
The phenomenon begins when hydrogen is ingested by a convective
zone driven by the energy flux from a unique helium-burning flash
in an EMP ($\feoh \lesssim - 2.5$) model star at 
the beginning of the TPAGB phase in low and intermediate-mass models.
     Panel (b) describes characteristics of the radiative
\nuc{13}{C}-burning model. The process repeats in every thermal
pulse cycle, and is made possible by the establishment, during
the third dredge-up phase, of a small region centered on the
dredge-up interface in which hydrogen and \nuc{12}{C} profiles
overlap, and a subsequent conversion of the \nuc{12}{C} in this
region into \nuc{13}{C} ``pocket.'' In this model, s-process
nucleosynthesis occurs during the interpulse phase in the
\nuc{13}{C} pocket.
     See the text for further details. 
}
\end{figure}

\begin{figure}
\epsscale{1}
\plotone{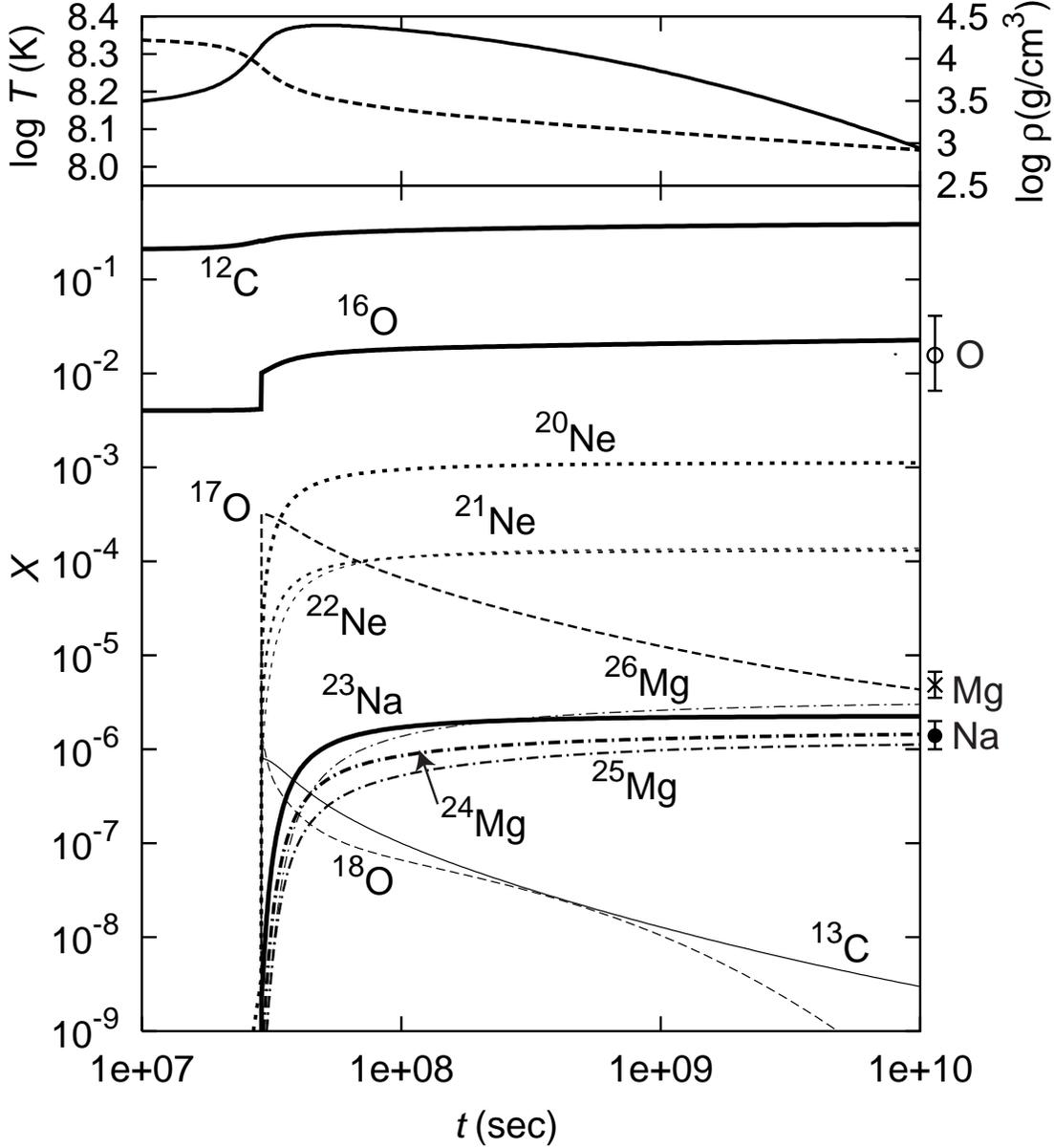}
\caption{
The lower panel shows the variation with time of isotopic
abundances in a helium-flash driven convective zone after
\nuc{13}{C} has been mixed into the zone. The initial
carbon isotope-abundance ratio in the zone is
$\nucm{13}{C} / \nucm{12}{C} = 0.001$ and the abundance
changes in the zone are driven by the release
of neutrons in the $^{13}$C$(\alpha,n)^{16}$O reaction.
     Instantaneous mixing of \nuc{13}{C} into the convective
zone is assumed to take place at the peak stage of helium burning.
     The top panel shows the variations in the temperature
(solid line) and in the density (broken line) during the helium
flash for model parameters taken from the $2 \msun$ star
by \citet{fuj00}.   
     Element abundances (relative to carbon) observed for \mmps\
are plotted at the right-hand side of the lower panel for
\nuc{}{O} (open circle), \nuc{}{Na} (filled circle),
and \nuc{}{Mg} (cross).
}
\end{figure}

\begin{figure}
\plotone{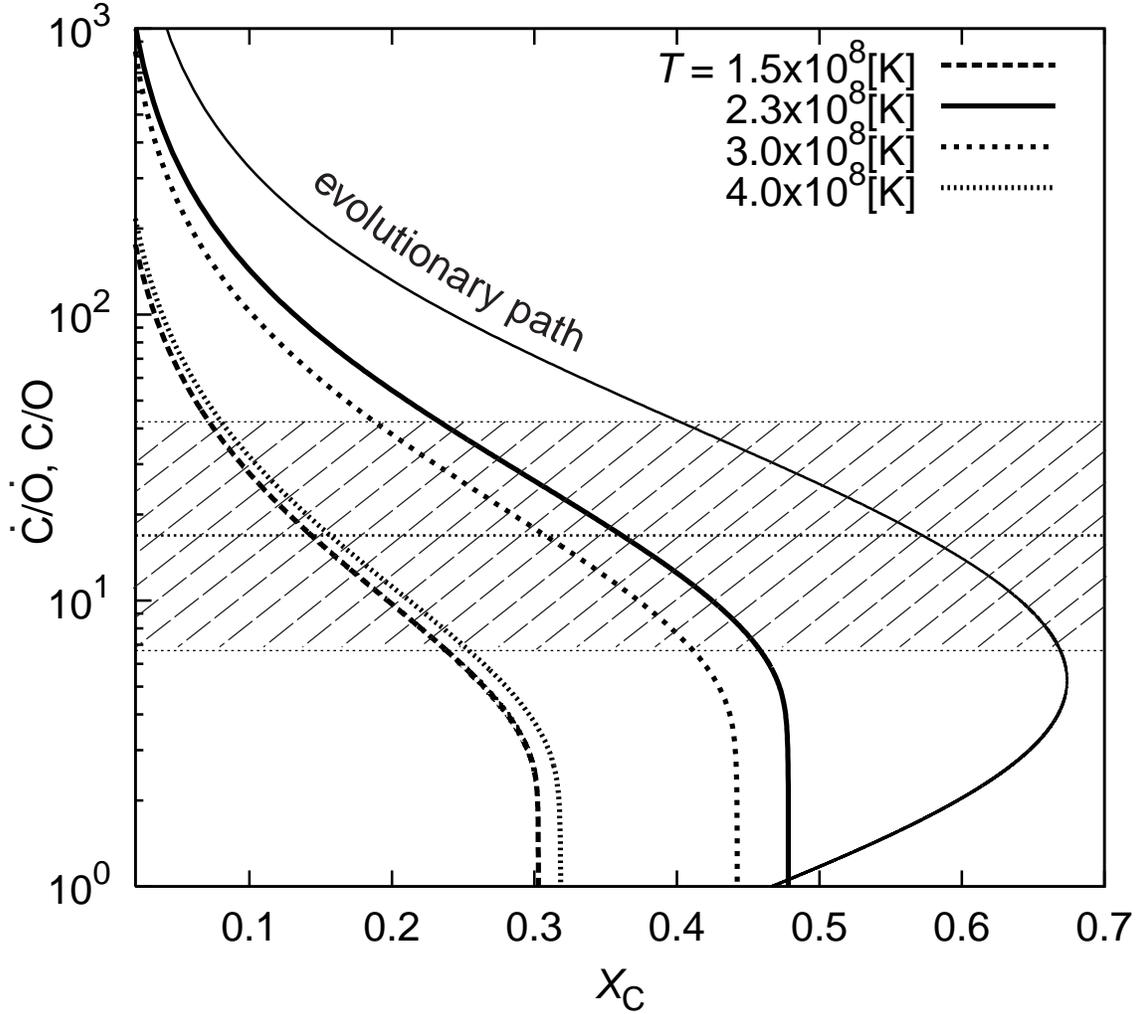}
\caption{
The ratio of the net production rate of carbon $\dot{\rm C}$ to the
production rate of oxygen $\dot{\rm O}$ due to the $3 \alpha$ and
\agco reactions as functions of the carbon abundance by mass
$X_{\rm C}$, for the temperatures listed at the top right corner.
The density has been set at $10^3\ \hbox{g cm}^{-3}$.  
     For a given value of $X_{\rm c}$, the helium abundance is set at $Y = 1 - X_{\rm C} (1 + \dot{\rm O} / \dot{\rm C})$, and there exists a limiting value of $X_{\rm C}$ corresponding to $\dot{\rm C} = 0$.  
   The thick solid line gives the ratio for the temperature
($T = 2.3 \times 10^8$K) at which the ratio is a maximum at
any given $X_{\rm C}$.  
     The thin solid line gives the evolutionary change in the
${\rm C} / {\rm O}$ abundance ratio starting from pure helium
(abundance by mass $Y = 1.0$) with $T = 2.3 \times 10^{8}$ K, for comparison's sake.
     The hatched area indicates the range in the ${\rm C} / {\rm O}$
abundance ratio estimated from the spectroscopic data for \mmps.
}
\end{figure}

\begin{figure}
\plotone{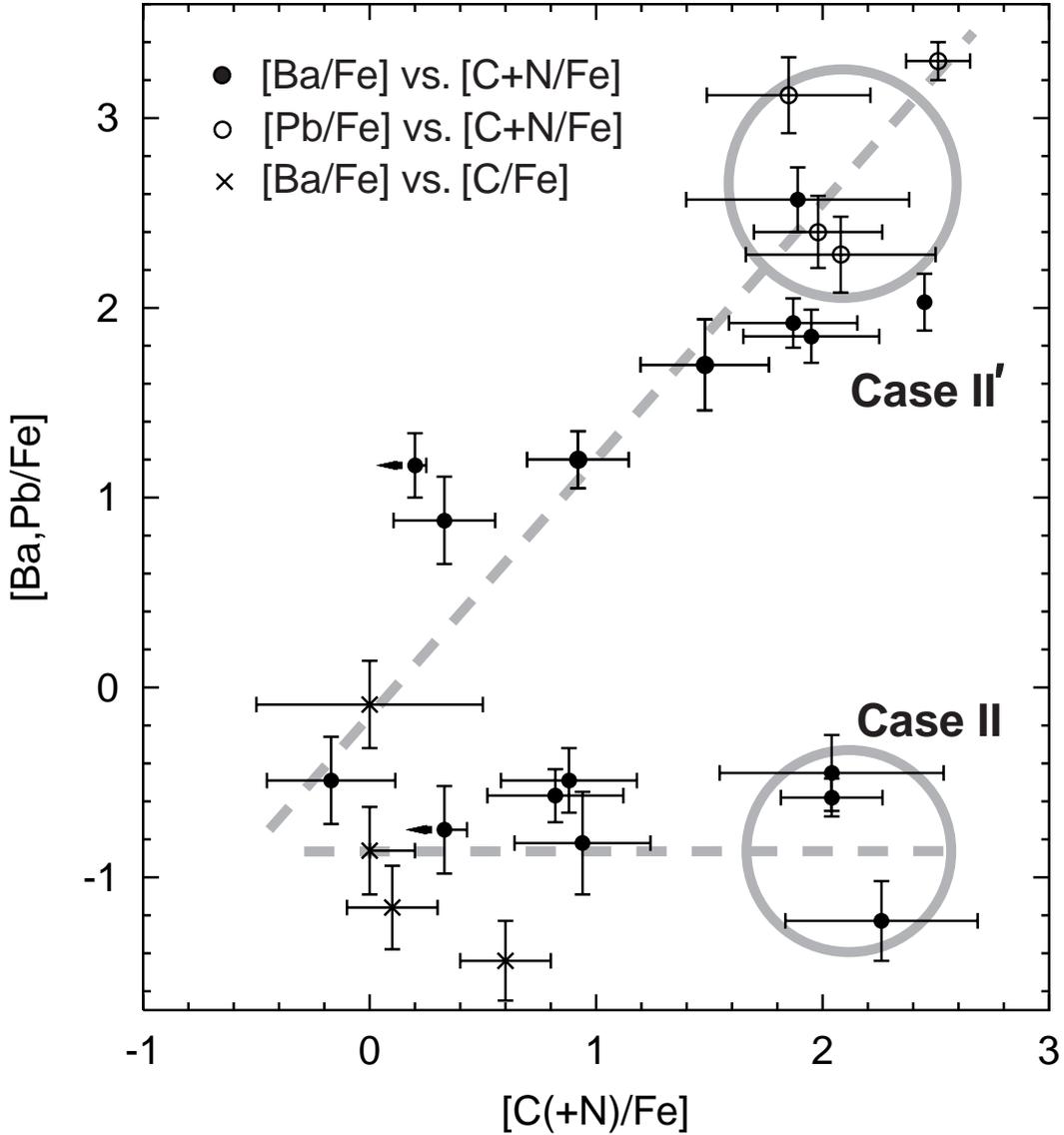}
\caption{
The distribution of EMP stars ($\feoh < -2.5$) with regard to
the enhancement of s-process elements versus the enhancement
of carbon and nitrogen. Literature sources are given in the text.
     The existence of two distinct branches defined by the
observed abundances is emphasized by a dashed line through 
each branch.
     The large open circles at the right hand ends of the two
branches indicate the abundances of CN and s-process elements
predicted, respectively, by Case II and Case II$^{\prime}$
evolutionary models.  
     Along each branch, the observed abundances are interpreted
to be the result of mixing, in a binary star system, between the
pristine envelope matter of a low-mass component and enriched
matter transferred from a more massive TPAGB star component.  
}
\end{figure}

\begin{figure}
\plotone{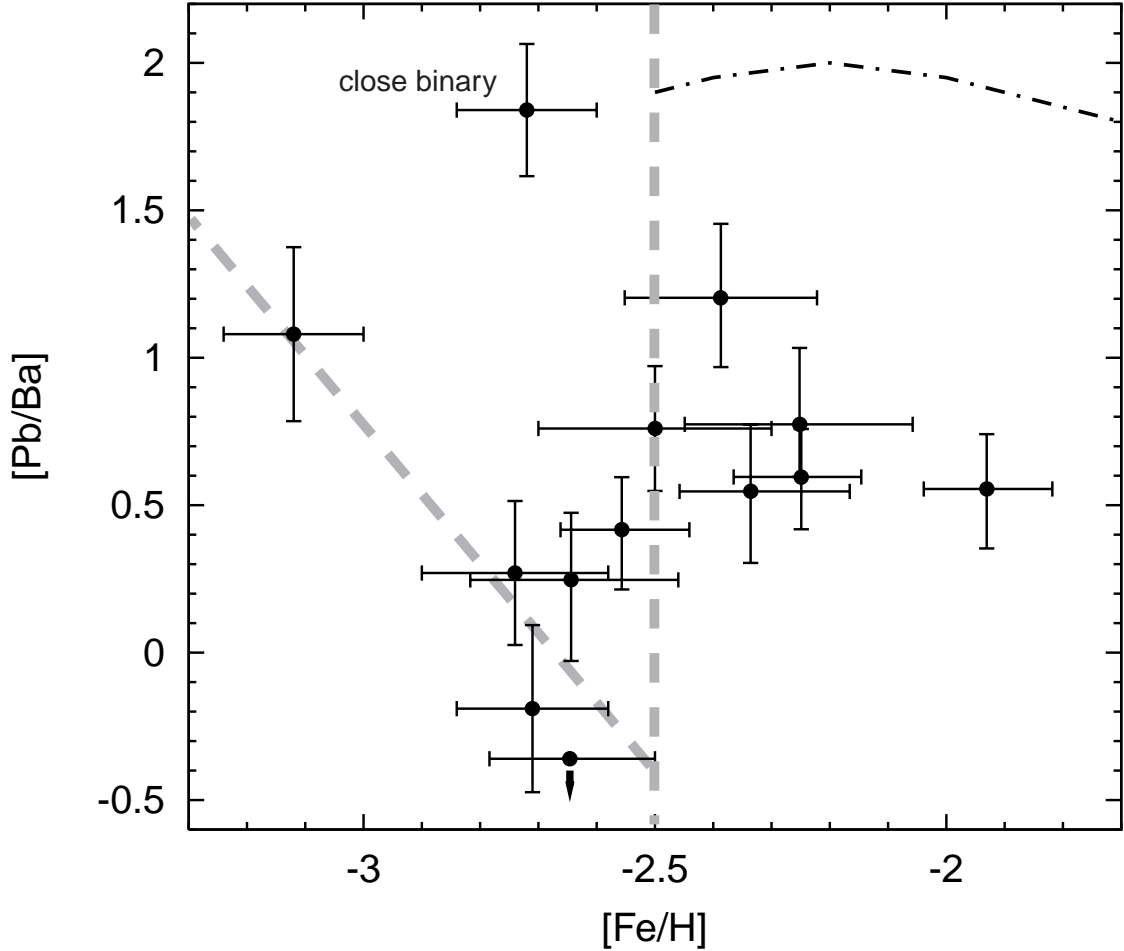}
\caption{
Variations of the ${\rm Pb}/{\rm Ba}$ ratio versus the metallicity.
Literature sources are given in the text.
     A break in the observed distribution is discernible near
$\feoh \simeq -2.5$, with ratios of $0.5 \sim 1.2$ on the larger
metallicity side, and ratios of $-0.5 \sim 0.5$ on the smaller
metallicity side.  
     This break may be interpreted as being due to a switch in
the mechanism of s-process nucleosynthesis from radiative
\nuc{13}{C} burning for Pop.~I and II stars to convective
\nuc{13}{C} burning for EMP stars.   
     The star HE 0024-2523, with $\feoh = -2.7$, is in a short period
binary ($P_{\rm orb}=3$ days), and the large 
$[{\rm Pb}/{\rm Ba}] \sim 1.84$ for this star may be attributed
to the effects on interior s-process nucleosynthesis of the common
envelope evolution that the binary star experienced while the primary
was in the TPAGB phase \citep{luc03}.  
     Excluding the point for the close binary star, one could argue
that, for $\feoh < -2.5$, the ${\rm Pb}/{\rm Ba}$ ratio
increases with decreasing metallicity. Such a trend is consistent
with the predictions of the He-FDDM convective \nuc{13}{C}-burning
model. The dash-dotted line in the region $\feoh > -2.5$ is the ratio
predicted by an s-process nucleosynthesis model \citep{bus99}
based on the radiative \nuc{13}{C}-burning model \citep{gali98}.
}
\end{figure}

\begin{figure}
\plotone{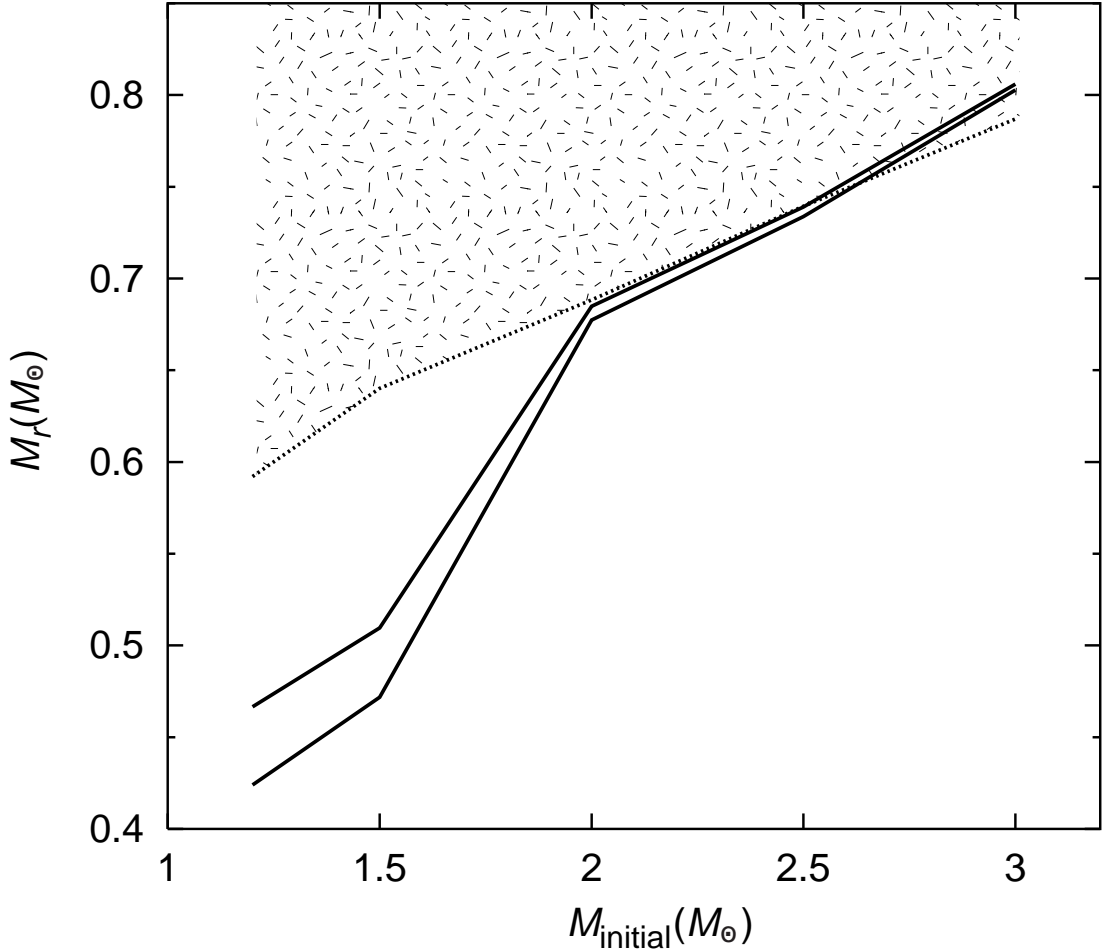}
\caption{
The deepest reach of surface convection during the second dredge-up
phase (denoted by the dotted line) and the locations of the top
and bottom of the He-FDDM convective zone when the top extends
through the base of the hydrogen profile (shown by the solid lines),
plotted as functions of the initial mass of Pop.~III model stars.  
     The hatched area identifies the zone into which surface
convection can carry accreted metals.  
     There exists a critical mass, $M_{\rm crit} \simeq
 2.5 \msun$, such that, for $M_{\rm initial}<M_{\rm crit}$,
the He-FDDM convective zone is confined to the metal-free zone,
while for $M_{\rm initial}>M_{\rm crit}$, it extends into or
lies entirely in the metal-polluted zone.  
     This penetration has been made possible by the fact that,
prior to the He-FDDM episode, model stars of mass
$M_{\rm initial}>M_{\rm crit}$ have experienced a second
dredge-up episode and hydrogen shell burning during the following
EAGB phase of evolution has moved the hydrogen-helium interface
into the metal-polluted zone.
}
\end{figure}

\begin{figure}
\plotone{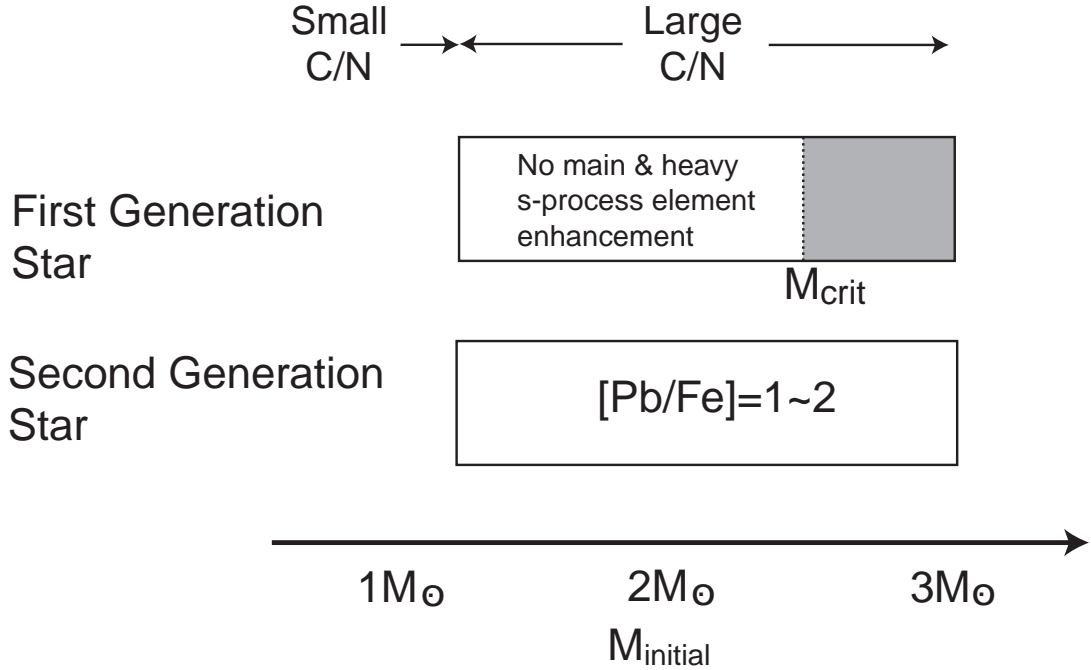}
\caption{
An illustration of the proposed binary scenario showing the
mass range of the primary component allowed by the currently
observed abundances of C, N, O, and Na in the spectrum of \mmps.  
     The upper box is for a first generation binary, having
acquired metals through accretion after birth, and the lower
box is for a second generation binary.  
     The allowed mass range corresponds to Case II$^{\prime}$
defined by FII00.  
     Predictions regarding s-process nucleosynthesis are
given in the boxes.  
     For a first generation binary, no enhancement in s-process
elements occurs if the initial primary mass is smaller than
$M_{\rm crit}\simeq 2.5$; an enhancement occurs if
the primary has an initial mass greater than $M_{\rm crit}$
(but less than $3 \msun$) and has accreted metals before
experiencing the He-FDDM event.
     For a second generation binary, an overabundance of heavy
s-process elements as large as $[{\rm Pb} / {\rm Fe}] = 1 \sim 2$
is expected.   
     A primary of mass less than that indicated in the figure
cannot produce a sufficiently large ${\rm C} / {\rm N}$ ratio
to account for the observations, while the He flash driven convective zone
in a primary more massive than indicated cannot reach the 
hydrogen-rich envelope. 
}
\end{figure}

\begin{figure}
\plotone{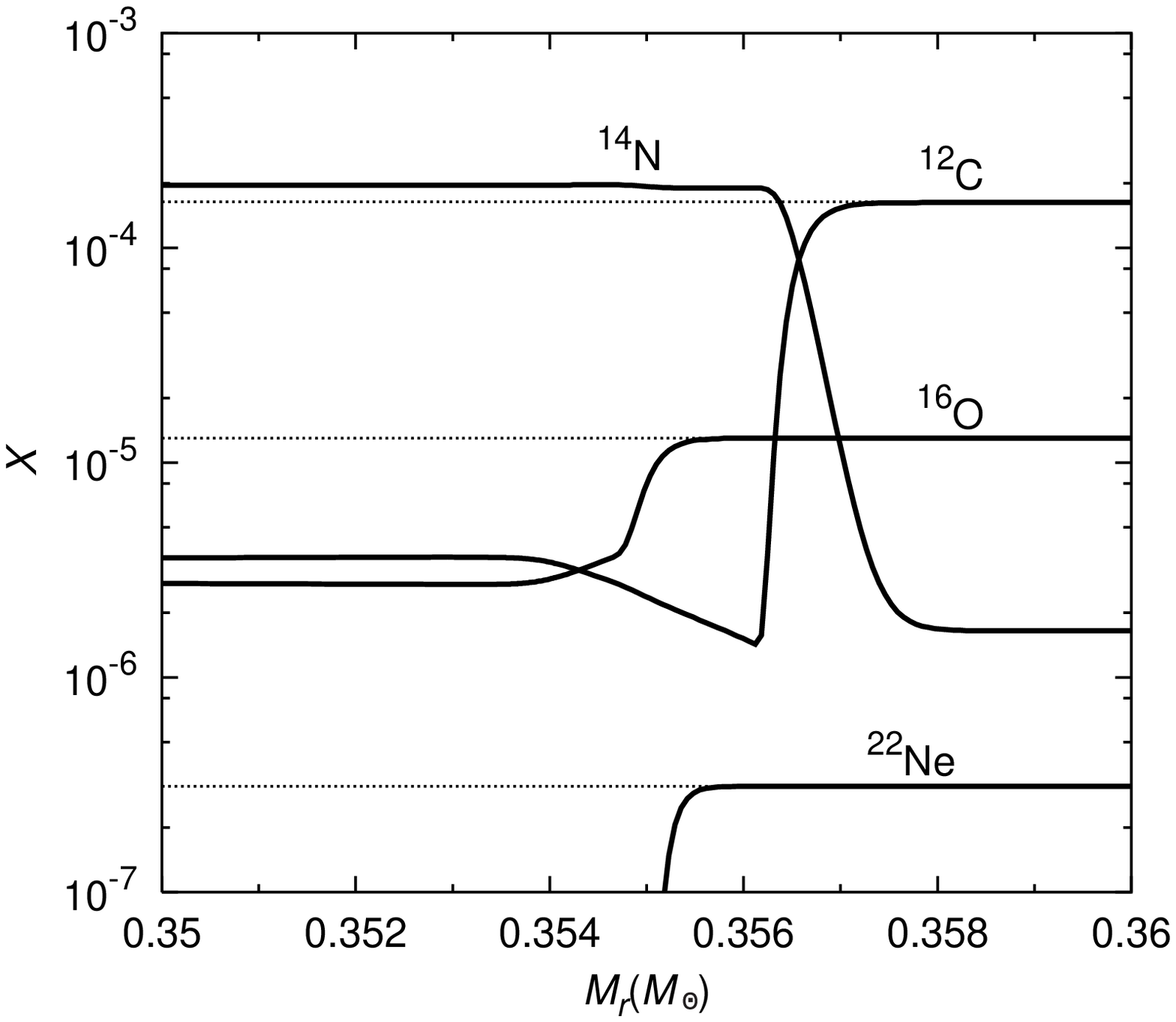}
\caption{
Abundance modifications due to the first dredge-up are
shown for a $0.8 \msun$ model star with the same C and O
abundances as \mmps.  
      The surface abundance of nitrogen increases to
$[{\rm N}/{\rm H}] = -2.8$, which falls within the observed
range of $[{\rm N}/{\rm H}] = -2.7 \sim -3.0$.   
     In order to see the change in the sodium abundance,
we add \nuc{22}{Ne}, but it cannot burn in the upper
part of the hydrogen-burning shell which is eventually mixed
into the envelope during the first dredge-up episode, so
no enrichment of sodium is obtained.   
}
\end{figure}
\end{document}